\documentclass[11pt]{article}
\usepackage{multicol}
\newtheorem{remark}{Remark}
\newtheorem{theorem}{Theorem}
\newtheorem{proposition}{Proposition}
\newtheorem{proof}{Proof}
\usepackage{graphicx}
\usepackage{dcolumn}
\usepackage{bbm}
\usepackage{bm}
\usepackage{geometry}
 \usepackage{cite}
\usepackage{braket}
\usepackage{lipsum}
\usepackage{mathtools}
\usepackage{lipsum}
\usepackage{bbold}
\usepackage{subcaption}

\usepackage{xcolor}
\usepackage[colorlinks=true,allcolors=blue]{hyperref}
\newcommand*{\SavedEqref}{}
\let\SavedEqref\eqref
\renewcommand*{\eqref}[1]{%
	\begingroup
	\hypersetup{
		linkcolor=linkequation,
		linkbordercolor=linkequation,
	}%
	\SavedEqref{#1}%
	\endgroup
}
\DeclareMathOperator{\Tr}{Tr}
\usepackage{algorithm}
\usepackage{algpseudocode}

\begin{document}
\title{Optimal probabilistic quantum control theory}

\author{ Randa Herzallah\thanks{\textcolor{blue}{Randa.Herzallah@warwick.ac.uk}}\hspace{0.2cm} and\hspace{0.1cm}Abdessamad Belfakir\thanks{\textcolor{blue}{abdobelfakir01@gmail.com}}\\Warwick Zeeman Institute of Mathematics, \\ The University of Warwick, Coventry, CV4 7AL, UK.}

\date{\vspace{-5ex}}
\maketitle
\begin{abstract}
There is a fundamental limit to what is knowable about atomic and molecular scale systems. This fuzziness is not always due to the act of measurement. Other contributing factors include system parameter uncertainty, functional uncertainty that originates from input functions, and sensors noises to mention a few. This indeterminism has led to major challenges in the development of accurate control methods for atomic scale systems. To address the probabilistic and uncertain nature of these systems, this work proposes a novel control framework that considers the representation of the system quantum states and the quantification of its physical properties following a probabilistic approach. Our framework is fully
probabilistic. It uses the Shannon relative entropy from information theory to design optimal
randomised controllers that can achieve a desired outcome of an atomic scale system. Several experiments are carried out to illustrate the applicability and effectiveness of the proposed approach.
\end{abstract}
 \section{Introduction }\label{sec1}
 In recent years, quantum control theory has been acknowledged as an enabling tool for the development of new quantum technology and quantum information theory applications \cite{Ramos,Krausz,Silberberg,Rego,Ohmori}. One of the main objectives in quantum control theory is to develop methods that can manipulate and control quantum systems. This has been achieved for the first time at the end of the last century \cite{Brumer,Kawashima,Weiner,Kosloff89,Jortner,Tannorand,Tannor} following the advent of laser femtosecond pulses. Consequent research studies investigated the development of optimal laser pulses as a control strategy to accomplish the required control objective \cite{Huang}. These studies were based on the variation of the laser pulse shape to optimize the outcome of an experiment, i.e. the result of  a predefined reaction product. This variation is repeated until the desired result is achieved \cite{Weiner2,Brown,Rabitz2003,Shapiro0,Brixner,Goswami}. 
 
 For the design of control strategy and methods, optimal control theory \cite{Werschnik,Zhu2000},  Lyapunov
 control approaches \cite{Mirrahimi,Kuang,Hou}, learning control algorithms \cite{DaoyiDong} and robust control methods \cite{Robust_1,Robust_2,Robust_3,Robust_4,Koswara} have
 been developed for the manipulation of quantum systems and the achievement of various control objectives. Among the aforementioned control design approaches, quantum optimal control is recognized as a powerful method for many complex quantum control tasks and has been successfully implemented for finding a control strategy for controlling molecules. In quantum optimal control, the control objective is usually casted as the optimisation of optimal cost function which is specified as the expected value of a target operator such as the projector onto a certain bound state or any other arbitrary operators \cite{Rice,Shapiro,Herek,Peirce98,RBZ98,RBZ982}. This optimisation of the cost function is commonly subject to some constraints including a penalty term on the energy radiation and the satisfaction of the Schr\"odinger equation by the wave function of the system. The solution of this constrained optimisation problem results into solving coupled non-linear  Schr\"odinger equations \cite{RBZ98,RBZ982}. 
 
Hence, several numerical methods have been introduced to optimise the cost functional including the conjugate gradient method \cite{Conju91} and the Krotov iteration method \cite{Kazakov}. However, it has been shown that most of these iterative methods are unreliable and computationally inefficient \cite{RBZ98}. Consequently, Zhu, Botina and Rabitz have proposed  the rapid
 monotonically convergent iteration method that solves the optimal control equations \cite{RBZ98,RBZ982}. It was proven that this algorithm recovers the Krotov iterative method as a particular case which was also tested for quantum optimal control of population and for quantum optimal control over
 the expectation value of a positive definite
 operator \cite{RBZ98,RBZ982}. In \cite{RBZ98}, the positivity of the target operator is shown to be mandatory for the assurance of the convergence of the cost functional.  Since then, several extensions to this algorithm have been constructed and practically applied  for the control of chemical reactions such as the feedback  and the resonant excitation strategies \cite{Nunes,Stensitzki}.
 
 Despite the success made so far in the area of quantum control, advances in this area are limited due to the high level of uncertainty, the effect of dissipation, and the probabilistic nature of atomic-scale physical systems.  A coherent control framework that can effectively address the aforementioned challenges is still lacking. This is mainly due to the fact that most of the existing advances in the field of quantum control are mostly based on designing deterministic controllers, thus they overlook the aforementioned challenges \cite{Brif,Londero}.  As such, this paper proposes  a fundamental new probabilistic feedback control framework for quantum systems. The proposed framework is fully probabilistic. It uses the Shannon relative entropy from information theory to design optimal randomised controllers \cite{Karny,RH_2011,RH_2013,RH_2015,RH_2018,RH_2020} that can achieve a desired closed loop performance of quantum systems under high level of uncertainty and stochasticity. To reemphasise, the derived control strategy under the proposed framework is fully probabilistic and it is based on the fact that the dynamics of a quantum system can be estimated using probability density functions (pdfs). We start by developing the general solution for the state space models of quantum systems whose state and output equations can be described by arbitrary probabilistic models. Using the vectorisation of the density operator obtained from the Liouville-von Neumann equation as the state equation and the measurements of the corresponding physical properties as the output equation, the solution to the proposed probabilistic control problem is then obtained and discussed in detail. Here, it will be shown that the distribution of the vectorised state of the density operator and output equations can be described using complex Gaussian distributions thus facilitating the analytic evaluation of the optimal randomised controller. For these quantum systems described by the Liouville-von Neumann equation, the optimal control solution is obtained as a feedback control law, which is linear in the state and whose gain matrix satisfies a Riccati equation. 
 
The proposed fully probabilistic control framework is different and more general to what has already been established in the literature. Its main characteristics are highlighted as follows: Firstly, it is a unified probabilistic control framework, all the components within this framework including the quantum controller, and quantum systems models are modelled using probabilistic models. This probabilistic characterisation of the individual components required to control quantum systems provides complete descriptions of their behaviours and depicts the inherent uncertainty in their dynamics. Secondly, it addresses the high level of uncertainty and the inherent probabilistic nature of atomic scale systems.  Thirdly, although an analytic solution of the proposed
method is only feasible for a vectorised quantum state whose time evolution is governed by linear and Gaussian distribution, the solution can be obtained in a closed form for any quantum system that can be described by arbitrary pdfs.
 
Contrary to existing deterministic control laws for quantum systems, the randomised controller obtained from the proposed probabilistic control framework is more explorative and provides complete information in the decision-making process, therefore is the natural solution to stochastic and uncertain atomic scale systems. It gives an optimal control solution in a feedback form that can be implemented either in real time by taking measurements from the controlled quantum system, or offline by feeding back estimated state values from the estimated probabilistic model of the time evolution of the system. For the online implementation though, quantum measurement backaction needs to be taken into consideration or otherwise a weak measurement procedure where only partial information on the measured observable is obtained will need to be followed. While it is interesting, this aspect of the study is beyond the scope of the current paper and thus further development will only focus on the establishment of the proposed probabilistic control framework.
 
The rest of this paper is organised as follows: Section (\ref{GeneralisedPD}) briefly recalls some preliminaries on the evolution of quantum systems and shows that their dynamics can be modelled using pdfs. It also states the objective of the considered control problem and provides its general solution. The proposed probabilistic control framework is then demonstrated in Section~(\ref{Solution}) on the bilinear representation of the vectorised time evolution of a quantum system described by the Liouville-von Neumann equation. In Section (\ref{app}), we particularly apply the proposed probabilistic control framework to the Lithium hybrid molecule $^7\text{Li}\hspace*{0.05cm}^2\text{H}$ and to particular spin systems interacting with external electric fields. Finally,  some conclusions are provided in Section (\ref{conclusion}).  

 \section{Fully Probabilistic Control for Quantum Systems}\label{GeneralisedPD}
 \subsection{Quantum System Description}\label{Quantum System Description}
 The time evolution of a quantum open system interacting with its environment can be described by the  Liouville von-Neumann equation,
 \begin{equation}\label{LVMS}
 	i \hbar\dfrac{d\rho(t)}{d t} =[H_0-\mu{u}(t),\rho(t)]+\mathcal{L}(\rho(t)), \hspace*{0,2cm} \rho(0)=\rho_0,
 \end{equation} 
 where $H_0$ is the system's free Hamiltonian, and $\hbar$ is the reduced Planck's constant. The system can be controlled using the electric field $u(t)$ where the interaction of the electric field with the quantum system can be described by the term $\mu{u}(t)$  and where $\mu$ is an operator related to the system  (e.g., electric
 dipole moment, polarizability). In addition, $\rho(t) \in \mathbf{C}^{l\times l} $ is the density operator which is a positive  hermitian operator with $\Tr(\rho(t))=1$, i.e. the eigenvalues of the density operator are interpreted as probabilities \cite{Neumann,Nemes}. The first term on the right hand side of Eq.(\ref{LVMS}) is the part that describes a closed system and it is called the Hamiltonian part. The second term is associated with the coupling of the system with the environment that is responsible of the dissipation. In the Lindblad  approach, the coupling term is provided by, 
 \begin{equation}\label{mathL}
 	\mathcal{L}(\rho(t))=i\hbar\sum_{s}(L_s\rho(t) L_s^\dagger-\dfrac{1}{2}\{L_{s}^\dagger L_s,\rho(t)\}),
 \end{equation}
 where $\{.\}$ stands for the anti-commutation operator, $L_s$ are the Lindblad operators and $s$ runs over all dissipation channels. 
 The operators $L_s$ are  defined in terms of the dissipative
 transition rates $\Gamma_{k\to j}$ from  the free Hamiltonian  eigenstate $\ket{k}$ to the eigenstate $\ket{j}$ as,
 \begin{equation}\label{L_s}
 	L_s=L_{j,k}=\sqrt{\Gamma_{k\to j}}\ket{j}\bra{k}.
 \end{equation}
 In this equation, it is assumed that $k$ takes a finite number of values, i.e., $\{\ket{k}, k = 0,..., l-1\}$  with $l$ being the number of the eigenvectors of the free Hamiltonian $H_0$.
In Appendix (\ref{app_A}), we show that the Liouville-von Neumann equation  (\ref{LVMS})  can be written   as,
 \begin{equation}\label{rho_elements}
 	\dfrac{d\rho_{n,m}(t)}{d t}=(-i\omega_{n,m}-\gamma_{n,m})\rho_{n,m}(t)+\sum_{k=0}^{l-1}\Gamma_{k\to n}\rho_{k,k}(t)\delta_{n,m}+i\dfrac{{u}(t)}{\hbar}\sum_{k=0}^{l-1}(\mu_{n,k}\rho_{k,m}(t)-\rho_{n,k}(t)\mu_{k,m}),
 \end{equation}
 where $\delta_{n,m}$ is the Kronecker symbol, $\{n,m\}=\{0,1,\dots,l-1\}$, $\mu_{k,n}:=\bra{k}\mu\ket{n}$ are the matrix elements of the operator $\mu$, $\omega_{n,m}:=\dfrac{E_n-E_m}{\hbar}$ are the Bohr frequencies, with $E_n$ being the energy eigenvalue of the free Hamiltonian $H_0$ associated with the eigenvector $\ket{n}$, and $\gamma_{n,m}$ is the total dephasing rate defined by,
 \begin{equation}\label{gamma_nm}
 	\gamma_{n,m}:=\dfrac{1}{2}\sum_{j=0}^{l-1}(\Gamma_{n\to j}+\Gamma_{m\to j}).
 \end{equation}

 
 The solution of the Liouville von-Neumann equation provides the information on the evolution of  physical properties of quantum open systems. The average value of an observable  $\hat{o}$ at a time instant $t$ is given by, 
 \begin{equation}
 	\braket{\hat{o}(t)} =\Tr(\hat{o} \rho(t)).
 \end{equation}
 \subsection{Control Objectives of the Quantum Control Problem} \label{ContObjQuantum}
Although open quantum systems can be characterised by the time evolution of the Liouville von-Neumann equation, this evolution may not be completely known and may be subject to uncertainty. This could be due to uncertainties from uncontrollable experimental parameter variations, and functional uncertainty that originates from input functions such as the variations in the manufacturing process for example. Considering the definition given in Appendix (\ref{app_B}) for the vectorised state vector, $x_t$ of the density matrix $\rho(t)$, the development in this paper will be based on the characterisation of its time evolution by a probability density function,
 \begin{equation}\label{pdfx1}
 	s(x_t|x_{t-1},u_{t-1}),
 \end{equation}
 where as previously defined, $u_{t}$ is the  electric field at time instant $t$. Note that because of the previously discussed uncertainties, the probabilistic description of the vectorised state vector of the density matrix as given in Eq.(\ref{pdfx1}) provides a complete specification of the present state $x_t$ as a
 function of the previous state, $x_{t-1}$ and previous control, $u_{t-1}$.
 
Similarly, observations of the observable, $\hat{o}$ are subject to different sources of uncertainties including sensors noises, and measurement uncertainties.  Thus, the probability density function of the observations provides the most complete specification of their values,
 \begin{equation}\label{pdfo1}
 	s(o_t|x_{t}).
 \end{equation}
 
The probabilistic description of the time evolution of the quantum system as given in Eq.(\ref{pdfx1}), and the observations as given in  Eq.(\ref{pdfo1}) is general and can be characterised by continuously monitoring its underlying stochastic evolution. It is not constrained by the linearity or Gaussian assumption of the stochastic evolution of the system. This characterisation is taken as a ready methodology in this paper, thus is not discussed further. Interested readers are referred to \cite{Mastriani_18,Emzir_17} on some of the available methodologies. 
 
 Following this formulation, the objective of the quantum control problem can be stated as follows: design a randomised controller, $c(u_{t-1}|x_{t-1})$ that minimises the Kullback-Leibler divergence between the
 joint pdf of the closed-loop description of the quantum system, $f(\mathcal{Z}(t, \mathcal{H}))$, and a predefined ideal joint pdf, $^If(\mathcal{Z}(t, \mathcal{H}))$,
 \begin{equation}\label{KLD}
 	\mathcal{D}(f||^I f)=\int f(\mathcal{Z}(t, \mathcal{H}))\ln\big(\dfrac{f(\mathcal{Z}(t, \mathcal{H}))}{^If(\mathcal{Z}(t, \mathcal{H}))}\big)d\mathcal{Z}(t, \mathcal{H}),
 \end{equation}
 where $\mathcal{Z}(t, \mathcal{H}) = \{x_t, \dots, x_\mathcal{H}, o_t, \dots, o_\mathcal{H}, u_{t-1}, \dots, u_\mathcal{H}\}$ is the closed-loop observed data sequence and $\mathcal{H} \le \infty$ is a given control horizon. The joint pdf of the closed-loop description of the system dynamics is the most complete probabilistic description of its behaviour. For the vectorised representation of the density matrix given in Eq.(\ref{pdfx1}), it can be evaluated using the chain rule \cite{Peterka81} as follows,
 \begin{equation}\label{JointDist}
 	f(\mathcal{Z}(t, \mathcal{H})) = \prod_{t=1}^\mathcal{H} s(x_t|x_{t-1},u_{t-1})s(o_t|x_{t})c(u_{t-1}|x_{t-1}),
 \end{equation}
 where the pdf $s(x_t|x_{t-1},u_{t-1})$ describes the dynamics of the vectorised density matrix, $s(o_t|x_{t})$ represents the dynamics of the observations $o_t$, and $c(u_{t-1}|x_{t-1})$ represents the pdf of the required randomised controller as mentioned earlier.
 
 Similarly, the ideal joint pdf of the closed-loop data can be factorised as follows,
 \begin{equation}\label{IdealJointDist}
 	^If(\mathcal{Z}(t, \mathcal{H})) = \prod_{t=1}^\mathcal{H} {^Is(x_t|x_{t-1},u_{t-1})} {^Is(o_t|x_{t})}{^Ic(u_{t-1}|x_{t-1})},
 \end{equation}
 where the pdf $^Is(x_t|x_{t-1},u_{t-1})$ describes the ideal distribution of the state vector of the vectorised density matrix, $x_t$, $^Is(o_t|x_{t})$ is the ideal distribution of the observations, and $^Ic(u_{t-1}|x_{t-1})$ represents the ideal pdf of the randomised controller. The definition of these ideal distributions is analogous to the target values in \cite{Demiralp93} that the system is required to achieve.
 
Given the definitions of the joint pdf of the closed-loop system and its ideal joint pdf as specified by Eqs.(\ref{JointDist}) and (\ref{IdealJointDist}), respectively, minimisation of Eq.(\ref{KLD}) can then be obtained recursively by introducing the following definition,

 	\begin{align}\label{OptPer}
 		-\ln(\gamma(x_{t-1})) &=  \underset{c(u_{t-1}|x_{t-1})}{\text{min}}\sum_{\tau=t}^\mathcal{H} \int f(\mathcal{Z}_t, \dots, \mathcal{Z}_\mathcal{H}|x_{t-1}) \ln \left( \dfrac{s(x_\tau|x_{\tau-1}, u_{\tau-1})s(o_\tau|x_\tau)c(u_{\tau-1}|x_{\tau-1})}{^Is(x_{\tau}|x_{\tau-1}, u_{\tau-1}) ^Is(o_\tau|x_\tau)^Ic(u_{\tau-1}|x_{\tau-1})} \right) \nonumber \\ &d(\mathcal{Z}_t, \dots, \mathcal{Z}_\mathcal{H}),
 	\end{align} 

 for arbitrary $\tau \in \{1, . . . , \mathcal{H}\}$. In Eq.(\ref{OptPer}), $-\ln(\gamma(x_{t-1})) $ specifies the expected minimum cost-to-go function, and $\mathcal{Z}_t = \{x_t,o_t,u_{t-1}\}$. Following the same procedure of classical physical systems \cite{Karny,RH_2011,RH_2013,RH_2015}, using the definition of the expected cost-to-go function given in Eq.(\ref{OptPer}), minimisation is
 then performed recursively to give the following recurrence functional equation,

 	\begin{align}\label{costtogo}
 		-\ln(\gamma(x_{t-1}))&=\underset{c(u_{t-1}|x_{t-1})}{\text{min}}\int s(x_t|x_{t-1}, u_{t-1})s(o_t|x_t)c(u_{t-1}|x_{t-1}) \nonumber \\ & \times \left [ \ln \left( \dfrac{s(x_t|x_{t-1}, u_{t-1})s(o_t|x_t)c(u_{t-1}|x_{t-1})}{^Is(x_t|x_{t-1}, u_{t-1}) ^Is(o_t|x_t)^Ic(u_{t-1}|x_{t-1})} \right) - \ln(\gamma(x_{t})) \right] d(x_t,o_t,u_{t-1}).
 	\end{align}

 \subsection{General Control Solution to the Quantum Control Problem}
 Following the representation given in Section (\ref{ContObjQuantum}) for the probabilistic state space models of open quantum systems described by the Liouville von-Neumann equation, the general solution for the optimal randomised controller that minimises the recurrence functional equation defined in
 Eq.(\ref{costtogo}) is given in the following proposition.
 \begin{proposition}\label{Prop1}	
 	The pdf of the optimal control law, $c(u_{t-1}|x_{t-1})$, that minimises the cost-to-go function (\ref{costtogo}) can be shown to be given by, 
 	\begin{equation}
 		\label{eq:Eqn4}
 		c(u_{t-1}|x_{t-1})=\frac{^Ic(u_{t-1}|x_{t-1}) \exp [-\beta(u_{t-1},x_{t-1})]}{\gamma(x_{t-1})},
 	\end{equation}
 	where
 	\begin{equation}\label{gamma2}
 		\gamma(x_{t-1}) = \int {^Ic(u_{t-1}|x_{t-1}) \exp[{-\beta(u_{t-1},x_{t-1})}] \mathrm{d} u_{t-1}},
 	\end{equation}
 	and

 		\begin{equation}\label{beta}
 			\beta(u_{t-1},x_{t-1}) = \int s(x_{t}|u_{t-1}, x_{t-1}) s(o_{t}|x_{t})\times \ln \bigg ( \frac{s(x_{t}|u_{t-1}, x_{t-1})s(o_{t}|x_{t}) }{^Is(x_{t}|u_{t-1}, x_{t-1}) ^I s(o_{t}|x_{t}) } \frac{1}{{\gamma}(x_{t})}\bigg) \mathrm{d} x_{t}\mathrm{d} o_{t}.
 		\end{equation}

 \end{proposition}
 \begin{proof}
 	The derivation of the above result can be obtained by evaluating the optimal cost-to-go function specified in Eq.(\ref{costtogo}).
 \end{proof}
To reemphasise, the randomised control solution given in Proposition (\ref{Prop1}) provides a general solution for quantum systems that are affected by various sources of uncertainties as explained earlier. Furthermore, it is not restricted by the Gaussian assumption of the pdfs of the quantum system states and observations or their ideal distributions. It provides the general solution for any arbitrary pdf. However, as will be seen in the following section, if all of the generative probabilistic models of the system dynamics, controller and ideal outcomes are Gaussian pdfs, an analytic form for the randomised controller can be obtained. 
 
 \section{Solution of the Quantum Control Problem for Gaussian pdfs}\label{Solution}
 This section will demonstrate the bilinear representation of the vectorised time evolution of a quantum system described by the Liouville-von Neumann equation (\ref{LVMS}) and discuss its characterisation with probabilistic models.   The theory developed in the previous section will then be applied here to derive the analytic solution of this bilinear state space model.
 \subsection{Bi-Linear State Space Model of the density matrix and measurements }\label{Bi-Linear Gaussian stochastic quantum systems}
 For quantum systems governed by the Liouville-von Neumann equation (\ref{LVMS}), that are also driven by external control field, $u_t$, the state vector of the vectorised density matrix is shown in Appendix (\ref{app_B}) to be given by the following bilinear equation \cite{wavepack1},
 \begin{align}\label{NLVN}
 	\dfrac{dx_{t}}{d t} &=(\tilde{A}+iu_t \tilde{N})x_{t}+i{q}u_t,\hspace{0.5cm} x_0=\tilde{x}, \nonumber\\
  &=\tilde{A} x_{t}+\tilde{B}(x_{t}) u_t ,\hspace{0.5cm} x_0=\tilde{x}, 
 \end{align}
 where $x_t\in \mathbf{C}^{n}$ is the vectorisation of the density operator $\rho(t)\in\mathbf{C}^{l\times l}$ with $n=l^2$, $\tilde{x}$ is the state of the system at time $t=0$, which is the initial condition of the differential equation (\ref{NLVN}), $\tilde{A}\in \mathbf{C}^{n\times n}$, 
 $\tilde{N}\in \mathbf{C}^{n\times n}$, $q\in \mathbf{C}^{n}$, and $\tilde{B}(x_t)= i( \tilde{N}x_{t}+q)$ \cite{wavepack1}. The elements of the matrices $\tilde{A}$, $ \tilde{N}$ and $q$ can be determined from the Liouville-von Neumann equation as explained in Appendix (\ref{app_B}). In contrast to linear systems, the coupling of  quantum systems with the external electric field induces a bilinear term  $\tilde{B}(x_t)$ that depends on the system's state $x_{t}$. Discretising the state space equation (\ref{NLVN}), the discrete time state space representation  can be obtained as follows,
 \begin{align}\label{StateQua0}
 	x_{t+1} = A x_t + B u_t + \tilde{w}_{t+1},
 \end{align}
 where,
 \begin{align}\label{A_t}
 	A &= e^{\tilde{A} \Delta t}, \\
 	\label{B_t}
 	B&= \bigg( \int_0^{\Delta t} e^{\tilde{A} \lambda} \tilde{B}(x_\lambda) \mathrm{d} \lambda\bigg), 
 \end{align}
 and where $\lambda = \Delta t - t$ with $\Delta t$ being the sampling period. We also have added the fictitious noise term $\tilde{w}_{t+1}$ to account for any uncertainty associated in the time evaluation of the density matrix as per the discussion in Section (\ref{ContObjQuantum}).  Equation (\ref{StateQua0}) can be equivalently written as,
 \begin{align}\label{StateQua}
 	x_{t} = A x_{t-1} + B u_{t-1} + \tilde{w}_{t}.
 \end{align}
 Similarly, the time evolution of the measurements of the operator $\hat{o}$ is given by,
 \begin{align}
 	o_t &=\Tr(\hat{o} \rho(t)) + \tilde{v_{t}},\\
 	&= \text{vec}(\hat{o}^T)^T \text{vec}(\rho(t))+ \tilde{v_{t}},\nonumber \\
 	\label{OutputQua0}
 	&= D x_t + \tilde{v_{t}},
 \end{align}
 where $D=\text{vec}(\hat{o}^T)^T \in \mathbf{C}^{(1\times n)}$ and $\tilde{v_{t}}$ is a fictitious noise that accounts for uncertainties caused by the measurement and environment. 
 
 Assuming that the noise $\tilde{w_{t}}$ affecting  Eq.(\ref{StateQua}) is a Gaussian noise, the pdf of the system state, $x_t$ can be considered as a complex normal pdf, 
 \begin{align}
 	\label{eq:eq4}
 	s(x_{t}\left| {x_{t - 1}},u_{t-1} \right.) &\sim \mathcal{N}_{\mathcal{C}}(\mu_{t},{\Gamma}), 
 \end{align}
 where,
 \begin{align}
 	\label{eq:relation1}
 	\mu_t& =E(x_t)=Ax_{t-1}+Bu_{t-1},\nonumber\\ 
 	\Gamma&= E((x_t-\mu_t)(x_t-\mu_t)^\dagger).
 \end{align}
 Here  $E(.)$ stands for the expected value,  $x_t^\dagger=\bar{x}^T$ is the conjugate transpose of $x_t$, and $x_t^T$ is the matrix transpose of $x_t$. The matrices $\mu_t$, and $\Gamma$  are respectively the mean and the covariance matrices. Also, note that the characterisation of the system state, $x_t$ by complex normal distribution is due to the fact that the system state vector is complex. The form of the complex normal distribution is recalled in Appendix (\ref{app_C}). 
 
 Similarly, assuming that the noise $\tilde{v_{t}}$ affecting  Eq.(\ref{OutputQua0}) is a Gaussian noise, the pdf associated with the measurement $o_t$ can be described by a standard normal distribution,
\begin{align}
	\label{eq:eq5}
	s({o_t}\left| {{x_{t}}} \right.) &\sim \mathcal{N}(o_{m_t},G),
\end{align}
where $o_{m_t}=Dx_{t}$ is the mean matrix and where,
\begin{align}
	G&= E((o_t-o_{m_t})(o_t-o_{m_t})^T)
\end{align}
is the covariance matrix of the  Gaussian distribution of the measurement vector, $o_t$. It is worth mentioning that the variable $o_t$ is real and for this reason its distribution should be taken as a standard normal distribution.
 
 \subsection{Optimal Control Law}
 In Section (\ref{Bi-Linear Gaussian stochastic quantum systems}), we have shown that the state vector of the vectorised density matrix described by Eq.(\ref{StateQua})  can be characterised by a complex normal distribution and the measurements  described by  Eq.(\ref{OutputQua0})  can be characterised by a  standard normal distribution, as stated in Eqs.  (\ref{eq:eq4}) and (\ref{eq:eq5}) respectively. Thus, the instantaneous joint pdf of the quantum system state, measurements  and control signal can be written as, 
 \begin{equation}\label{gpdf2}
	f(x_t,o_t,u_{t-1}|x_{t-1})=c(u_{t-1}|x_{t-1})\mathcal{N}(o_{m_t},G)\mathcal{N}_{\mathcal{C}}(\mu_{t},{\Gamma}),
\end{equation}
 where $c(u_{t-1}|x_{t-1})$ is the randomised optimal control to be derived. Accordingly, the ideal pdfs of the system state, the measurements and the controller are taken to be Gaussians as follows,
 \begin{align}
	\label{eq:eq1}
	{}^Is({x_{t}}\left| {{x_{t - 1}}},u_{t-1} \right.) &\sim \mathcal{N}_{\mathcal{C}}(x_r,{\Gamma_r}), \\
	\label{eq:eq2}
	{}^Is({o_t}\left| {{x_{t}}} \right.) &\sim \mathcal{N}(o_{r},G_r),\\
	\label{eq:eq3}
	{}^Ic({u_{t-1}}\left| {x_{t - 1}} \right.) &\sim \mathcal{N}(u_r,\Omega).
\end{align}
 Thus, the instantaneous ideal joint pdf can now be written as, 
 \begin{equation}\label{igpdf2}
 	^If(x_t,o_t,u_{t-1})=\mathcal{N}(u_r,\Omega)\mathcal{N}(o_{r},G_r)\mathcal{N}_{\mathcal{C}}(x_r,{\Gamma_r}).
 \end{equation}
 The objective here is to find the distribution of the controller $c({u_{t-1}}\left| {x_{t - 1}} \right.) $  provided in Eq.(\ref{gpdf2}) that minimises the Kullback-Leibler distance between the joint distribution (\ref{gpdf2})  and the predefined ideal one (\ref{igpdf2}). The minimisation of the Kullback-Leibler distance as proposed in the current paper is the analogous objective functional to the deterministic one in \cite{Demiralp93} for obtaining optimal control laws for stochastic systems described by probabilistic models as given in Eqs.(\ref{eq:eq4}) and (\ref{eq:eq5}) . 
 
 The derivation of the controller distribution appearing in Eq.(\ref{gpdf2}) that minimises the Kullback-Leibler divergence between the pdf (\ref{gpdf2}) and the ideal one (\ref{igpdf2}) is based on the evaluation of the optimal performance index, $-\ln(\gamma(x_t))$. The form of this controller will be stated shortly, but first the form of the optimal performance index is provided in the following theorem. 

 	\begin{theorem}\label{Theo1}
 		By substituting the ideal distribution of the system dynamics (\ref{eq:eq1}), (\ref{eq:eq2}), the ideal distribution of the controller (\ref{eq:eq3}), and the real distribution of the system state (\ref{eq:eq4}) and measurements (\ref{eq:eq5}) into Eq.(\ref{gamma2}), the performance index can be shown to be given by,
 			\begin{align}
 			- \ln \left( {\gamma \left( {{x_{t-1}}} \right)} \right) &={x}_{t-1}^\dagger{M_{t-1}}{x_{t-1}}+{P}_{t-1}{x_{t-1}}+\bar{P}_{t-1}{\bar{x}_{t-1}} + {\omega_{t-1}}, \label{gamma_theo}
 		\end{align}
 		where, 
 		\begin{align}
 			\label{Mt}
 			{M_{t - 1}} &=A^{\dagger}\bigg(C-CBF_{t}^{-1}B^\dagger{C}\bigg)A,
 		\end{align}
 		\begin{align}
 			\label{Pt}
 			P_{t-1}&=\bigg((u_r^\dagger\Omega^{-1}+JB)F_{t}^{-1}B^{\dagger}C-J\bigg)A
 		\end{align}
 		and where,
 		\begin{align}
 			\label{Const}
 			\omega_{t-1}&=u_r^\dagger\Omega^{-1}u_r+x_r^{\dagger}\Gamma_r^{-1}{x}_r+o_r^{\dagger}G_r^{-1}o_r+\omega_{t}+I- \ln \left(\dfrac{|F_{t}^{-1}|}{|\Omega|}\right).	
 		\end{align}	
 		In additions, in Eqs. (\ref{Mt})-(\ref{Const}) the following definitions were used,
 		\begin{align}\label{parameters}
 			&	C= \Gamma_r^{-1}+D^{\dagger}G_r^{-1}D+M_{t},
 			\nonumber\\
 		&	J=	x_r^{\dagger}\Gamma_r^{-1}+o_r^\dagger{G}_r^{-1}D-P_t,
 		 			\nonumber\\	& F_t={\Omega^{-1}}+B^\dagger{C}B  \nonumber\\ I=&
 		 			\ln \bigg( \dfrac{|\Gamma_r|}{|\Gamma|} \dfrac{|G_r|}{|G|}\bigg)-\Tr(G({G}^{-1}-{G}_r^{-1}))-	\Tr(\Gamma({\Gamma}^{-1}-{\Gamma}_r^{-1}-D^{\dagger}G_r^{-1}{D}-M_t)).
 		\end{align}
 	\end{theorem}

 \begin{proof}
 	The proof of this theorem can be obtained by using backward induction as shown in Appendix (\ref{AppD}). We start by computing the coefficient $\beta(u_{t-1},x_{t-1})$ using the definition (\ref{beta}), which will then be followed by the computation of $\gamma(x_{t-1})$ using the definition (\ref{gamma2}). This will yield the assumed form in the theorem. 
 \end{proof}
 \begin{remark}
 	From  Eq.(\ref{gamma_theo}), it can be clearly seen that the optimal cost-to-go function, $-\ln(x_t)$ maps the complex vector $x_t\in\mathbf{C}^n$  to a real number, since $M$ is an hermitian operator. This preserves the important necessity that cost functions are designed to be ordered and that their values can be compared in a consistent ordering manner. 
 \end{remark} 
 
 Following the form of $\gamma(x_{t-1})$ specified in Theorem (\ref{Theo1}), it is now straight forward to derive the pdf of the optimal control law that minimises the Kullback Leibler divergence between the actual pdf (\ref{gpdf2}) and the ideal one (\ref{igpdf2}). It can be obtained by substituting Eqs.(\ref{gamma2}) and (\ref{beta}) in Eq.(\ref{eq:Eqn4}) yielding the following theorem.
 \begin{theorem}\label{Theo2}
 	The distribution of the optimal control law that  minimises the Kullback-Leibler distance between (\ref{gpdf2}) and (\ref{igpdf2}) is Gaussian distribution given by, 
 	\begin{equation}\label{optimalc}
 		c(u_{t-1}|x_{t-1})\sim \mathcal{N}(v_{t-1},F_{t}^{-1}).
 	\end{equation}
 	where 
 	\begin{equation}\label{optimalc2}
 		v_{t-1}=-F_{t}^{-1}\bigg(B^\dagger{C}Ax_{t-1}-\Omega^{-1}u_{r}-B^\dagger{J}^\dagger\bigg).
 	\end{equation}
 	with $F_t$, $C$ and $J$ being provided in Eq.(\ref{parameters}). In addition, $v_{t-1}$ is the mean of the Gaussian distribution of the optimal control law and $F_{t}^{-1}$ is its variance.
 \end{theorem}
 
 \begin{proof}The proof of this theorem is given in Appendix (\ref{AppE}). 
 \end{proof}
 
 \begin{remark} 
Although the derived probabilistic controller (\ref{optimalc}) for the assumed probability distributions maintains the standard form of linear quadratic controllers, it is more exploratory due to its probabilistic nature. For the implementation of this controller to real world systems, either the mean of the randomised controller can be taken to be the optimal control input, or control inputs should be
ideally sampled from the obtained pdf of the randomised controller. Sampling control inputs from the obtained pdf  however results in slightly worse control quality, but randomisation makes the controller more explorative.
\end{remark}
 
 \subsection{Implementation Procedure of the Randomised Controller}
 The step by step implementation of the fully probabilistic control problem of quantum systems where the characterising pdfs are assumed to be Gaussians is provided as pseudocode in Algorithm (\ref{alg:cap}). The execution of this algorithm facilitates the evaluation of the optimal electric field that drives the quantum system from its initial state to a predefined desired one. 

 	\begin{algorithm}[H]
 		\caption{Fully probabilistic control of quantum systems\hspace{4cm}}\label{alg:cap} 
 		\begin{algorithmic}[1]
 			\State  $\text{Evaluate the operator} \hspace{0.1cm} D\hspace{0.1cm} \text{associated with the target operator}\hspace{0.1cm} \hat{o}\text{;}$
 			\State  $\text{Compute the matrices} \hspace{0.1cm} \tilde{A}\hspace{0.1cm} \text{and} \hspace{0.1cm} \tilde{N}\hspace{0.1cm} \text{from Eq. (\ref{NLVN}), hence evaluate}\hspace{0.1cm} A\hspace{0.1cm} \hspace{0.1cm} \text{from  Eq.(\ref{A_t})} \text{;}$
 			\State  $\text{Determine the predefined  desired value ${o}_{r}$, and  the predefined  desired state ${x}_{r}$;}$
 			\State  $\text{Specify the initial state $x_0$  and then calculate the initial value of the measurement}$ $\text{ state $o_0\gets Dx_0$;}$
 			\State  $\text{Provide the covariance of the ideal distribution of the observation vector, $G_r$, the ideal distribution}$ $ \text{ of the state vector, $\Gamma_r$, and the covariance of the controller, $\Omega$;}$
 			\State  $\text{Initialise:}$ $t\gets0$, $M_0 \gets \text{rand}$, $P_0 \gets \text{rand}\text{;}$
 			\While{$t \ne \mathcal{H}$}
 			\State  $\text{Evaluate} \hspace{0.1cm} B\hspace{0.1cm} \hspace{0.1cm} \text{from  Eq.(\ref{B_t})} \text{;}$
 			\State $\text{Calculate the steady state solutions of} \hspace{0.1cm} M_t\hspace{0.1cm} \text{and}\hspace{0.1cm}  P_t\hspace{0.1cm}\text{following the formulas provided in  Eqs.(\ref{Mt}),}$ $\text{ and (\ref{Pt}) respectively};$
 			\State $\text{Use}\hspace{0.2cm} M_t,  \hspace{0.2cm}\text{and}\hspace{0.2cm} P_t\hspace{0.2cm} \text{ to compute the mean of the optimal control input, $v_{t-1}$  following  Eq.(\ref{optimalc2}) given in}$ $\text{ Theorem (\ref{Theo2});} $ 
 			\State  $\text{Set:}$ $u_{t-1}\gets v_{t-1}$; 
 			\State $\text{Using the obtained control input from the previous step, evaluate \hspace{0.1cm}  $x_t $ according to Eq.(\ref{StateQua}),}$ $\text{${x}_{t}\gets Ax_{t-1}+B(x_{t-1}) u_{t-1} +\tilde{w}_{t} $} $
 			\State $\text{Following  Eq.(\ref{OutputQua0}), evaluate $o_t$ to find the measurement state at time instant $t$, $o_t \gets Dx_t+\tilde{v_{t}}$ ;}$
 			\State $t\gets t+1\text{;}$	
 			\EndWhile
 		\end{algorithmic}
 	\end{algorithm}
Algorithm (\ref{alg:cap}) will be applied in the next section to control certain quantum physical systems.
 \section{Application to particular physical systems}\label{app}
 \subsection{Generalities on Morse potential}
 Following \cite{Morse,Landau}, the Morse potential  is a solvable model that very well describes the vibrations inside diatomic molecules \cite{Jensen,Pauling}. For a diatomic molecule, the Morse potential is governed by,
 \begin{equation}\label{MorseHamilt}
 	V_M(r)=D_{0}(e^{-2\alpha (r-r_{\text{eq}})}-2e^{-\alpha (r-r_{\text{eq}})}),
 \end{equation}
 where $r$  is the distance between the two atoms and $r_{\text{eq}}$ is the corresponding equilibrium position \cite{Morse,Landau}, while $D_{0}$ and $\alpha$ are related to the depth and the width of the potential, respectively. The parameter $D_0$ is related to the molecule properties by $\nu=\sqrt{\frac{8m_{r}D_{0}}{\alpha^{2}\hbar^{2}}}$, where $\nu$ is a parameter corresponding to the spectroscopic constants of  the molecule and $m_{r}$  is the reduced mass of the oscillating atoms. The Morse potential is a solvable model and its energy eigenvalues are given by,
 \begin{equation}
 	\label{spect}
 E_n=-\dfrac{\hbar^2\alpha^2}{2m_r}\big(n-p\big)^2,
  \end{equation}
where $n=\{0,1,\dots [p]\}$ with $p=\dfrac{\nu-1}{2}$, and $[.]$ denotes the integer part operation. Hence, the number of the system's free Hamiltonian defined in Section (\ref{Quantum System Description}) is now $l=[p]+1$. Without loss of generality, in the following we omit the effect of the environment and we consider that the molecule only interacts with the electric field $u_t$. This means that the Linblad operators appearing in Eq.(\ref{LVMS}) are now $L_s=0$ for any $s$. The interaction between the system and the electric field is described within the semi-classical approximation through the electric dipole $\mu\equiv\mu(r)=\mu_0 r\text{e}^{-r/r^{*}}$ where  $ \mu_0 $ and $r^*$ are parameters related to the molecule \cite{Herzberg}. The matrix elements of $\mu$  can be computed by,
\begin{equation}\label{electricdipol}
	\mu_{n,m}=\bra{n}\mu\ket{m}=\int_{-\infty}^{\infty} {\mu}(r)\psi_{n}^{\nu}(r)  \psi_{m}^{\nu}(r){d}{r},
\end{equation}
where $\{n,m\}=\{0,\dots,[p]\}$ and $\psi_{n}^{\nu}(r)$ are the energy eigenfunctions associated with $E_n$. They are given by, 
\begin{equation}\label{psiMorse}
		\psi_{n}^{\nu}(r)=\text{N}_{n}e^{-\frac{y}{2}}y^{j}L_{n}^{2j}(y),
\end{equation}
where we have used the change of variable  $y=\nu e^{-\alpha (r-r_{\text{eq}})}$, and where $L_{n}^{2j}(y)$ are the Laguerre polynomials. In addition, $2j=\nu-2n-1$, and
$\text{N}_{n} $ is the normalization factor governed by, 
\begin{equation}\label{Norm}
	\text{N}_{n}=\sqrt{\frac{\alpha(\nu-2n-1)\Gamma(n+1)}{\Gamma(\nu-n)}},
\end{equation}
where $\Gamma$ is the gamma function\cite{Morse}.
 \subsection{Application to the Lithium hybrid molecule}
 Here we consider the  Lithium hybrid molecule whose Morse parameters can be found in \cite{Herzberg}. Their values are given by    $D_0=2.45090 \hspace{0.1cm}\text{eV}$, $r_{\text{eq}}=2.379 \hspace{0.1cm}\text{{\AA}
} $ and $\nu\approx 6.1346$. Hence, in this case $p=2.5673$, which means that the quantum number $n$, appearing in Eq.(\ref{spect}) of the Morse potential associated with the Lithium hybrid molecule takes three values $0,1,2$.   As explained in Section (\ref{LVMS}), the interaction  between the electric field  and the physical system is realised through the electric dipole operator $\mu$ which is now given by $\mu=\mu(r)=\mu_0 r\text{e}^{-r/r^{*}}$ where  $ \mu_0 $ and $r^*$ are parameters related to the molecule \cite{Herzberg}. Their values for Lithium hybrid molecule are given by $\mu_0\approx5.8677\hspace{0.2cm} \text{Debye}$ and $r^*\approx1.595\text{{\AA}}$. By omitting the effect of the environment, the time evolution of the density operator (\ref{rho_elements})  can be governed by,
	\begin{equation}
\label{rho_elements_NoEnv}
\dfrac{d\rho_{n,m}(t)}{d t}=-i\omega_{n,m}\rho_{n,m}(t)+i\dfrac{u(t)}{\hbar}\sum_{k=0}^{l-1=2}(\mu_{n,k}\rho_{k,m}(t)-\rho_{n,k}(t)\mu_{k,m}),
\end{equation} 
where in this example, $\{n,m\}=\{0,1,2\}$.\\
 Although any target operator can be considered, here we consider the following Gaussian operator given in the position representation $\{\ket{r}\}$ as follows,
\begin{equation}\label{targetoper}
	\hat{o}=\hat{o}(r)=\dfrac{\gamma_0}{\sqrt{\pi}}\text{e}^{-\gamma_0^2(r-r^{'})^2},
\end{equation}
where $\gamma_0=25$ and $r^{'}=2.4871 \hspace{0.1cm}\text{{\AA}}$.
The matrix representation of the target operator (\ref{targetoper}) can then be obtained as follows,
\begin{equation}
	o_{i,j}=\int_{-\infty}^{\infty} {o}(r)\psi_{i}^{\nu}(r)  \psi_{j}^{\nu}(r){d}{r},\hspace{0.3cm}\text{where}\hspace{0.3cm}\{i,j\}=\{0,1,2\}.
\end{equation} 
Therefore, following the vectorisation method described in Eq.(\ref{x_t_vet}), the operator $D$ that appears in the measurement equation (\ref{OutputQua0}) can be easily constructed. It is given by,
\begin{equation}
	D=[o_{0,0}\hspace{0.2cm} o_{1,1}\hspace{0.2cm}o_{2,2}\hspace{0.2cm}o_{0,1}\hspace{0.2cm}o_{0,2}\hspace{0.2cm}o_{1,0}\hspace{0.2cm}o_{2,0}\hspace{0.2cm}o_{1,2}\hspace{0.2cm}o_{2,1}].
\end{equation}
 Using the definition of vectorisation of the density matrix provided in  Eq.(\ref{x_t_vet}) along with  Eq.(\ref{rho_elements_NoEnv}) and   Eq.(\ref{electricdipol}), the parameters of the state equation defined in  Eq.(\ref{NLVN}) can be easily obtained as follows, 
\begin{equation}
	\tilde{A}=\text{diag}[0 \hspace{0.2cm}0\hspace{0.2cm}0\hspace{0.2cm} -i\omega_{0,1}\hspace{0.2cm}  -i\omega_{0,2}\hspace{0.2cm} -i\omega_{1,0}\hspace{0.2cm} -i\omega_{2,0}\hspace{0.2cm} -i\omega_{1,2}\hspace{0.2cm}-i\omega_{2,1}],
\end{equation}
\begin{equation}
	\tilde{N}=\dfrac{1}{\hbar}\left(
	\begin{array}{ccccccccc}
		0 & 0 & 0 &-\mu_{0,1} & -\mu_{2,0} & \mu_{0,1} &\mu_{0,2} & 0 & 0 \\
		0 & 0 & 0 &\mu_{1,0} & 0 & -\mu_{0,1} & 0 &- \mu_{2,1} & \mu_{1,2}\\
		0 & 0 & 0 & 0 & \mu_{2,0} & 0 & -\mu_{0,2} &\mu_{2,1} &- \mu_{1,2} \\
		-\mu_{0,1} & \mu_{0,1} & 0 & \mu_{0,0}-\mu_{1,1} & -\mu_{2,1} & 0 & 0 & 0 & \mu_{0,2} \\
		-\mu_{0,2} & 0 &\mu_{0,2} & -\mu_{1,2} & \mu_{0,0}-\mu_{2,2}  & 0 & 0 & \mu_{0,1}  & 0 \\
		\mu_{1,0}  & -\mu_{1,0} & 0 & 0 & 0 &-\mu_{0,0}+\mu_{1,1} & \mu_{1,2} & -\mu_{2,0} & 0 \\
		\mu_{2,0} & 0 & -\mu_{2,0} & 0 & 0 & \mu_{2,1} & -\mu_{0,0}+\mu_{2,2} & 0 & -\mu_{1,0} \\
		0 & -\mu_{1,2}  & \mu_{1,2}  & 0 & \mu_{1,0}  & -\mu_{0,2}  & 0 & \mu_{1,1} -\mu_{2,2} & 0 \\
		0 & \mu_{2,1}  & -\mu_{2,1} & \mu_{2,0} & 0 & 0 & -\mu_{0,1} & 0 & -\mu_{1,1}+\mu_{2,2}\\
	\end{array}
	\right),\end{equation}
where $\tilde{A}$ is a diagonal $9\times9$-matrix. The shifting vector $x_e$ appearing in Eq.(\ref{a_til}) is $x_e={x}_0$ with ${x}_0$ being the vectorisation of the density matrix at $t=0$, considered here to be the ground state. Furthermore, Using  Eqs.(\ref{A_t}) and (\ref{B_t}), we can easily find the matrix elements of  $A$ and $B$ operators determining the evolution of the vectorised state $x_t$ at each time step $t$, as governed in Eq.(\ref{StateQua0}).  Taking $ \Delta{t}=0.0167$, $G_r=0.000000014$,  $\Gamma_r=10$ and $\Omega=0.28950$, we evaluate the matrices $M_t$ and $P_t$ defined in Eqs. (\ref{Mt}) and (\ref{Pt}), respectively  at each instant of time as discussed in Algorithm (\ref{alg:cap}). Then, we use these matrices to evaluate the control signal, $u_{t-1}$ following Eq.(\ref{optimalc2}). The obtained electric field is used in Eq.(\ref{StateQua}) to find the evolution of the state vector $x_t$ and the corresponding observation can be determined using Eq.(\ref{OutputQua0}). We repeat these steps until the measurement output, $o_t$ becomes as close as possible to the  predefined desired value ${o}_{r}$, which is taken to be equal to $1$ in this example, i.e., ${o}_{r}=1$.

\begin{figure}[h!]
	\begin{subfigure}{.5\textwidth}
		\centering
		\includegraphics[width=.8\linewidth]{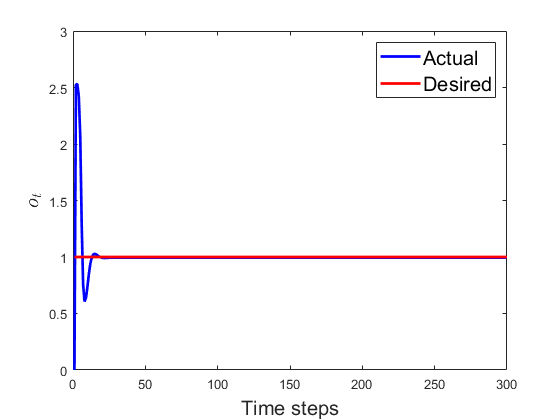}  
	\caption{}
		\label{fig:sub-first}
	\end{subfigure}
	\begin{subfigure}{.5\textwidth}
		\centering
		\includegraphics[width=.8\linewidth]{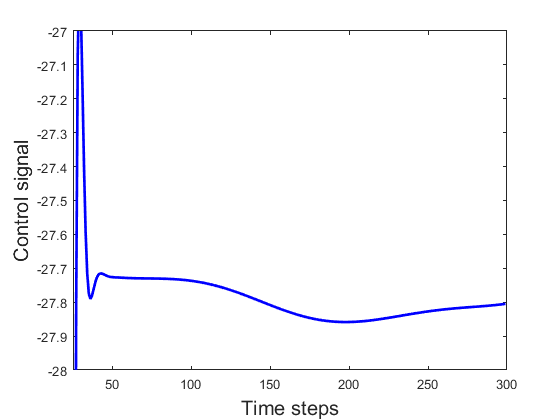}  
		\caption{}
		\label{fig:sub-second}
	\end{subfigure}
	\caption{(\ref{fig:sub-first} ): The blue curve represents the time evolution of the average value of the target Gaussian operator (\ref{targetoper}) for the Morse potential associated with the Lithium hydride molecule $^7\text{Li}\hspace*{0.05cm}^2\text{H}$  under the effect of the control signal. The red curve is for the desired value of the Gaussian operator, i.e. $o_r=1$. (\ref{fig:sub-second}): The time evolution of   the control signal,  $u_t$ responsible of the transformation of the system state to the predefined target state $x_r$ associated with $o_r=1$. In both figures atomic unites are used.}
	\label{figure_1}
\end{figure}

  Figure (\ref{fig:sub-first}) shows the behaviour of the time evolution of the average value of the target operator (\ref{targetoper}). The corresponding control signal that allows the achievement of the control objective is shown in Figure (\ref{fig:sub-second}). 
   In all figures we used the atomic units.   In the examined example, it can be clearly seen that under the effect of the electric field the average value of $o_t$ tends to the target value $o_r$ which means that the  control objective is achieved. This demonstrates the effectiveness of the method introduced in this paper. 
\subsection{Spin systems}
The introduced probabilistic control strategy will be implemented here to manipulate the transition probabilities of a number of examples of spin-$j$ systems. A spin-$j$ system can be described by,
\begin{equation}
	\ket{\psi}=\sum_{m=-j}^{j}c_m\ket{j,m},
\end{equation}
where $c_m$ are complex coefficients that satisfy $\sum_{m=-j}^{j}|c_m|^2=1$, and the ensemble $\{\ket{j,m}\equiv\ket{m}, m=-j,\dots,j\}$ are  the common eigenstates of the operators $J^2=J_1^2+J_2^2+J_3^2$ and $J_3$. In addition, $J_k$ with  $(k = 1,2,3)$ are some observables satisfying the angular momentum
commutation relations $[J_1,J_2]=i\hbar\epsilon_{123}J_3$ (with $\epsilon_{123}$ being the Levi-civita symbol). We consider that the spin-$j$ system interacts with an external electric field and its evolution can be totally described by the following Hamiltonian,
\begin{equation}
	H= H_0+ H_u(t),
\end{equation}
where $H_0$ is the system's free Hamiltonian and $H_u(t)$ is the operator describing the interaction between the system and the electric field $u_t\equiv{u}(t)$. The control objective here is to transfer the system from an initial state, $\ket{\psi_i}$ to a predefined final state $\ket{\psi_f}$ through its interaction with the optimised electric field. Hence, the problem can be seen as a maximisation of the fidelity between the initial and the final states; $\ket{\psi_i}$ and $\ket{\psi_f}$, respectively. This means that the target operator here is nothing but the projector $\Pi_f=\ket{\psi_f}\bra{\psi_f}$. With this objective, the time evolution of the observation $o_t$ given in Eq.(\ref{OutputQua0}) becomes the time evolution of the population of the final state, $\ket{\psi_f}$  of the considered spin system. Thus, the value of the measurement  $o_t$ will converge to  $1$ when the system successfully transits to the final state. This means that the desired value of the observation, in this case, is $o_r=1$. Indeed, the problem of maximisation of the fidelity between the initial and the final states has been solved using different approaches such as the rapidly convergent algorithm in \cite{RBZ982} and the method investigated in \cite{spin}. 

Next, we particularly consider spin-1/2 system, i.e., $j=\dfrac{1}{2}$ and spin-1 system, i.e., $j=1$.
\subsubsection{spin-$\dfrac{1}{2}$ system}
\label{spin1_2}
The probabilistic control method discussed in Section(\ref{ContObjQuantum}) will be applied here to a spin-$\dfrac{1}{2}$ system interacting with an electric field $u_t$. According to \cite{spin}, the dimensionless Hamiltonian describing the interaction between the system and the electric filed is given by,
\begin{equation}\label{58_}
	H=H_0+H_u(t)=\dfrac{1}{2}\sigma_3+\dfrac{1}{2}(\sigma_1+\sigma_2)u_t,
\end{equation} 
where  $\sigma_1$, $\sigma_2$ and $\sigma_3$ are the Pauli matrices given in the basis $\{\ket{-}\equiv\ket{\dfrac{1}{2},-\dfrac{1}{2}},\ket{+}\equiv\ket{\dfrac{1}{2},\dfrac{1}{2}}\}$ as,
\begin{equation}\label{PauliMat}
	\sigma_1	=\left(
	\begin{array}{cc}
		0&1\\
		1 & 0 \\
	\end{array}
	\right),\hspace{0.4cm}	\sigma_2=\left(
	\begin{array}{cc}
		0 &-i\\
		i & 0 \\
	\end{array}
	\right),\hspace{0.4cm}\sigma_3	=\left(
	\begin{array}{cc}
		1 &0\\
		0 & -1\\
	\end{array}
	\right).
\end{equation}
We consider that the system is initially prepared in the state $\ket{\psi_i}=\ket{-}$, and the objective is to transfer it to the state $\ket{\psi_f}=\ket{+}$ through its interaction with the  optimised electric field $u_t$ that is designed to achieve that objective. The calculation of this electric field at each time step is explained in Algorithm (\ref{alg:cap}). In particular, using  Eq.(\ref{OutputQua0}), the target operator is evaluated to be  $D=[0\hspace{0.2cm}1 \hspace{0.2cm}0\hspace{0.2cm}0]$, which is nothing but $D=(\text{vect}(\Pi_+))^T$, where $\Pi_+=\ket{+}\bra{+}$. Next, the values of the matrices $\tilde{A}$ and $\tilde{N}$ are computed as explained in Appendix (\ref{spinn_1_2}) followed by the calculation of the operators $A$ and $B$ at each time instant using Eqs.(\ref{A_t}) and (\ref{B_t}) and taking $\Delta{t}=0.00071439$. By considering those matrices and taking $G_r=0.0000001$, $\Omega=0.0001$ and $\Gamma_r=1000$, the operators $M_t$ and $P_t$ are then computed at each time step according to Eqs.(\ref{Mt}) and (\ref{Pt}), respectively. Those are used to compute the optimal electric field $u_{t-1}$ using Eq.(\ref{optimalc2}) which in turn is used to evaluate the evolution of the system using Eq.(\ref{StateQua}). These steps are repeated until the output, $o_t$ becomes as close as possible to the desired one, $o_r=1$.  
\begin{figure}[h!]
	\begin{subfigure}{.5\textwidth}
		\centering
		\includegraphics[width=.8\linewidth]{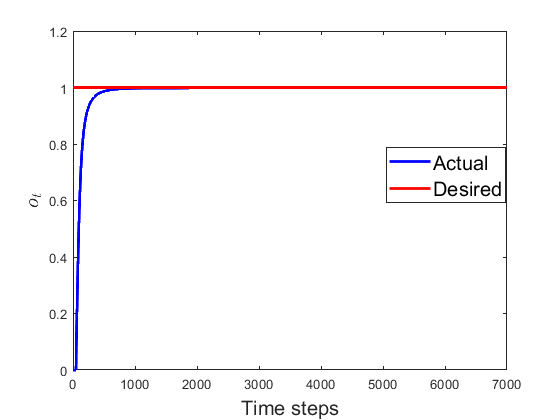}  
		\caption{}
		\label{spin3}
	\end{subfigure}
	\begin{subfigure}{.5\textwidth}
		\centering
		\includegraphics[width=.8\linewidth]{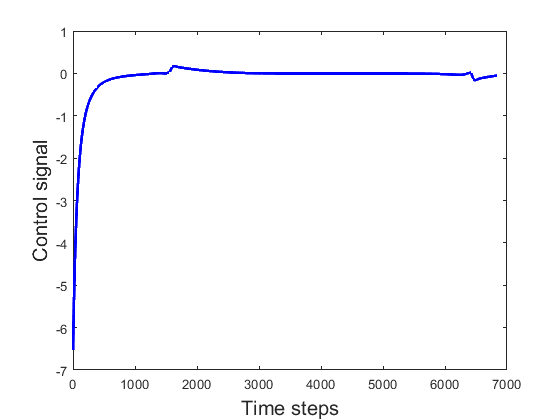}  
		\caption{}
		\label{spin1_4}
	\end{subfigure}
	\caption{(\ref{spin3} ): The blue curve represents the time evolution of the population of the $\ket{+}$ state, $\rho_{11}(t)$ of the considered spin-1/2 system. The red curve is the desired value ${o}_{r}=1$. (\ref{spin1_4}): The time evolution of  the control signal,  $u_t$ responsible of achieving the control objective.}
	\label{figure_2_2}
\end{figure}
Figure (\ref{spin3}), shows the time evolution of the population of the  $\ket{+}$ state, $\rho_{11}(t)$ of the considered spin-$\dfrac{1}{2}$ system interacting with the electric field. We can clearly see that the population  $\rho_{11}$ has reached the predefined target value $o_r$ in few time steps. This demonstrates the effectiveness of the proposed control method. Figure (\ref{spin1_4}) shows the time evolution of the optimal electric field responsible of transferring the system from its initial state to the desired one.
\subsubsection{Spin-$1$}
\begin{figure}[h!]
	\begin{subfigure}{.5\textwidth}
		\centering
		\includegraphics[width=.8\linewidth]{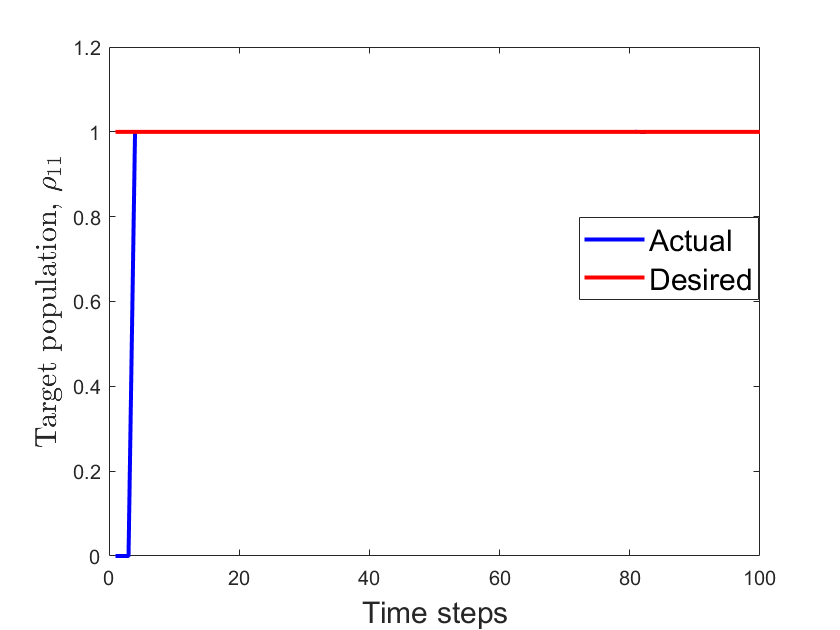}  
		\caption{}
		\label{figure5}
	\end{subfigure}
	\begin{subfigure}{.5\textwidth}
		\centering
		\includegraphics[width=.8\linewidth]{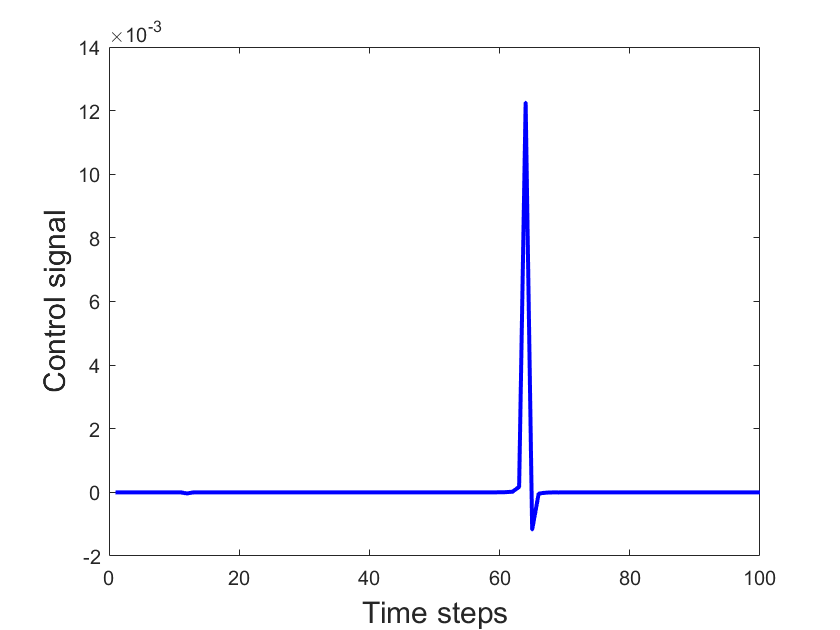}  
		\caption{}
		\label{figure6}
	\end{subfigure}\\
	\begin{subfigure}{.5\textwidth}
		\centering
		\includegraphics[width=.8\linewidth]{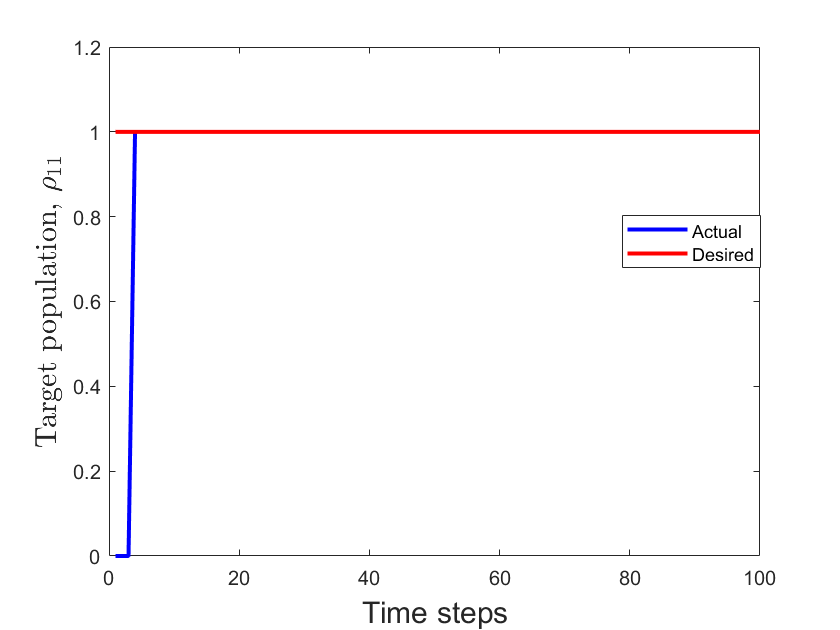}  
		\caption{}
		\label{figure7}
	\end{subfigure}
	\begin{subfigure}{.5\textwidth}
		\centering
		\includegraphics[width=.8\linewidth]{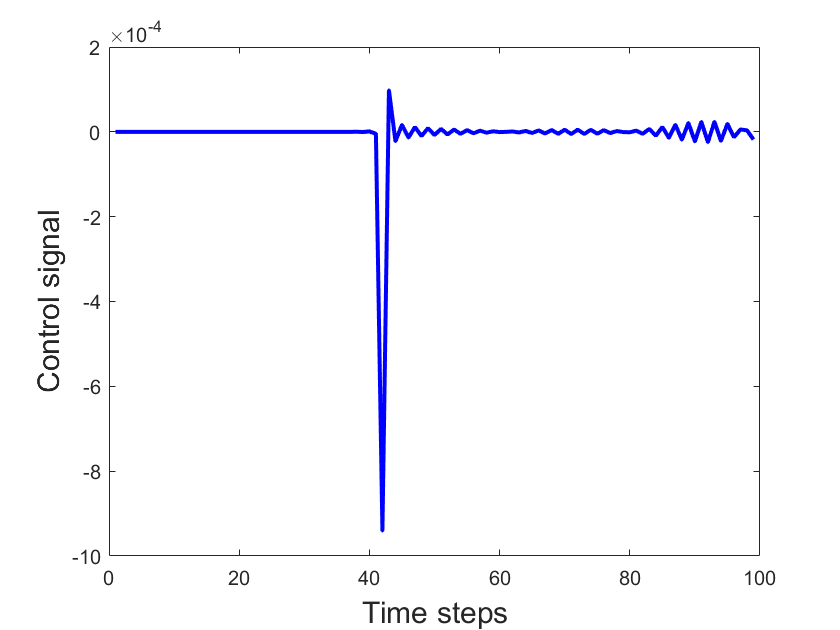}  
		\caption{}
		\label{figure8}
	\end{subfigure}
	\caption{(\ref{figure5} ): The blue curve represents the time evolution of the population of the $\ket{1,1}$ state, $\rho_{11}(t)$ of the considered spin-1 system initially prepared in the state $\ket{1,-1}$. The red curve is the desired value ${o}_{r}=1$. (\ref{figure6}): The time evolution of  the control signal,  $u_t$ responsible for achieving the control objective. (\ref{figure7} ):  The blue curve represents the time evolution of the population of the $\ket{1,1}$ state, $\rho_{11}(t)$ of the considered spin-1 system initially prepared in the state $\ket{1,0}$. The red curve is the desired value ${o}_{r}=1$. (\ref{figure8}): The time evolution of  the control signal,  $u_t$ responsible for achieving the control objective.}
	\label{figure_3}
\end{figure}

To further demonstrate the effectiveness of the probabilistic approach introduced in this work, it is applied to a spin-$1$ system   interacting with an electric field, $u_t$. We consider that the interaction can be described by the following dimensionless Hamiltonian,
\begin{equation}
	\label{Hamil1}
	H=H_0+H_u(t)=\left(\begin{array}{ccc}
		\dfrac{3}{2} & 0 & 0  \\
		0 & 1 & 0  \\
		0 & 0&0   \\
	\end{array}\right)+\left(\begin{array}{ccc}
		0 & 0 & 1  \\
		0 & 0 & 1  \\
		1 & 1&0   \\
	\end{array}\right) u(t).
\end{equation}  
The objective here is to transfer the system from an initial state to the desired state $\ket{1,1}$, which means that the target operator is $D=(\text{vect}(\Pi_1))^T$ where $\Pi_1=\ket{1,1}\bra{1,1}$. Appendix (\ref{spinn_1}) demonstrates how to evaluate the matrices $\tilde{A}$ and $\tilde{N}$ associated with the spin-$1$ system whose evolution is described by Eq.(\ref{Hamil1}), and gives their values. The first objective is to transfer the system from the ground state $\ket{1,-1}$ to state $\ket{1,1}$ and the second objective is to transfer the system from the state $\ket{1,0}$ to the  state $\ket{1,1}$. Taking $G_r=0.000000001$,
$\Omega=0.11$ and $\Gamma_r=10000$ for the first objective and $G_r=0.000000005$,
$\Omega=0.1$ and $\Gamma_r=10000$ for the second objective, and using the steps explained in Algorithm (\ref{alg:cap}), the operators $A$, $B$, $M_t$, $P_t$ and  then the optimal control $u_{t-1}$ are determined at each time step, $t$. This electric field, $u_{t-1}$ is then used to evaluate the evolution of the system and these steps are repeated until the measurement output, $o_t$ becomes as close as possible to the  predefined desired value, ${o}_{r}$ which is evaluated to be equal to $1$ in both experiments. Figure (\ref{figure5}) shows  the time evolution of the population of the  $\ket{1,1}$ state, $\rho_{11}(t)$ of the considered spin-$1$ system initially prepared in the state $\ket{1,-1}$ while  Figure (\ref{figure7} ) shows  the time evolution of the population of the  $\ket{1,1}$ state for the spin-$1$ system initially prepared in the state $\ket{1,0}$. Figures (\ref{figure6}) and (\ref{figure8}) show the time evolution of the associated optimal control signals, responsible for achieving the first and second control objectives respectively. It can be clearly seen, in both experiments, that  the designed optimal control input, $u_{t-1}$ was successful in transitioning, in a few time steps, the system state from the initial states to the desired final state, $\ket{1,1}$, demonstrating the effectiveness and the simplicity of the proposed control method. 

Finally, in all experiments, it can be clearly seen that once converged the time evolutions of the populations of the systems demonstrated in this section were maintained through the designed controller at their desired values for longer time steps which is a very important achievement. 
 \section{Final comments}\label{conclusion}
 In this paper we have introduced a new probabilistic strategy to control physical systems at the atomic and molecular scales. The main particularity of our approach compared to the existing ones is that it is fully probabilistic and considers the probabilistic nature of the dynamics of quantum systems  due to  different sources  of uncertainty and stochasity. Due to these uncertainties, we have shown that the  dynamics of quantum systems can be characterised by probability density functions. For quantum systems described by the Liouville-von Neumann equation, the pdfs of their dynamics represented by the vectorisation of their corresponding density operator and the measurements of an observable are then shown to be Gaussian under the assumption that the noise affecting them is white Gaussian noise. Thus,  Gaussian probability density functions are associated to the state of the system  and the measurement of its physical properties. Furthermore, using the minimisation of the Kullback-Leibler divergence between the joint pdf of the actual state, measurements and controller and a predefined ideal joint pdf, we have provided the form of the control law that transfers the system from its initial state to the desired one. Moreover, we have applied the proposed approach to  the Lithium hybrid and spin systems in interaction with an external electric field   and shown that the state of  the outcome associated with particular target operators can be controlled using the derived optimal control signal. 
\section*{Acknowledgements:} This work was supported by the EPSRC grant EP/V048074/1.
 
 \newpage
 \section*{Appendices}
 
 \appendix
 \renewcommand{\theequation}{A.\arabic{equation}}
 \setcounter{equation}{0} 
 
 	\section{Evolution of quantum open systems and state space model}\label{app_A}
 The time evolution of a quantum open system interacting with its environment can be described by the  Liouville von-Neumann equation (\ref{LVMS}).
In Dirac notations, the free Hamiltonian $H_0$ can be written as, \begin{equation}\label{Ham}
	H_0=\sum_{n=0}^{l-1}E_n\ket{n}\bra{n},
\end{equation}
where $E_n$ are the energy eigenvalues of $H_0$. In the vector space spanned by the energy eigenvectors,
the density matrix can be written as, 
\begin{equation}\label{rh}
	\rho(t)=\sum_{k,j=0}^{l-1}\rho_{kj}\ket{k}\bra{j},
\end{equation}	
where $\rho_{kj}$ are the matrix elements of $\rho(t)$. Hence, it follows that,
\begin{equation}\label{9}
	[H_0,\rho(t)]=\sum_{k,j=0}^{l-1}(E_k-E_j)\rho_{kj}\ket{k}\bra{j}.
\end{equation}
Similarly, the commutator between the operator $\mu$ and the density matrix $\rho(t)$ is given by,
\begin{equation}\label{mu_rho}
	[\mu,\rho(t)]=\sum_{k,j,m=0}^{l-1}\mu_{kj}\rho_{jm}\ket{k}\bra{m}-\sum_{k,j,n=0}^{l-1}\rho_{nk}\mu_{kj}\ket{n}\bra{j},
\end{equation}
where $\mu_{k,j}=\bra{k}\mu\ket{j}$. Now, we calculate the elements of the open system operator $\mathcal{L}(\rho(t))$ appearing in Eq.(\ref{mathL}). The first term in Eq.(\ref{mathL}) is governed by,
 \begin{equation}\label{cp2}
	\sum_{s}L_{s}\rho(t) L_{s}^\dagger=	\sum_{j,k=0}^{l-1}L_{jk}\rho(t) L_{jk}^\dagger= \sum_{j,k=0}^{l-1}\Gamma_{k\to j}\rho_{kk}\ket{j}\bra{j},
\end{equation}
where we used Eq.(\ref{L_s}). The anti-commutator appearing in Eq.(\ref{mathL}) can be given by,
\begin{align}\label{cp3}
	\dfrac{1}{2}	\sum_{s} \{L_{s}^\dagger L_{s},\rho(t)\}= \dfrac{1}{2}	\sum_{j,n,m=0}^{l-1}\big(\Gamma_{n\to j}+\Gamma_{m\to j}\big)\rho_{nm}\ket{n}\bra{m}.
\end{align}
By substituting Eqs.(\ref{9}), (\ref{mu_rho}), (\ref{cp2}) and (\ref{cp3}) in Eq.(\ref{LVMS}) we find,
\begin{align}\label{cp4}
	i\hbar\dfrac{d}{dt}\sum_{k,j=0}^{l-1}\rho_{kj}\ket{k}\bra{j}
	&=\sum_{k,j=0}^{l-1}(E_k-E_j)\rho_{kj}\ket{k}\bra{j}+u(t)\bigg[-\sum_{k,j,m=0}^{l-1}\mu_{kj}\rho_{jm}\ket{k}\bra{m}+\sum_{k,j,n=0}^{l-1}\mu_{kj}\rho_{nk}\ket{n}\bra{j}\bigg]\nonumber \\&
	+i\hbar\sum_{j,k=0}^{l-1}\Gamma_{k\to j}\rho_{kk}\ket{j}\bra{j} -i\hbar\dfrac{1}{2}	\sum_{j,n,m=0}^{l-1}\big(\Gamma_{n\to j}+\Gamma_{m\to j}\big)\rho_{nm}\ket{n}\bra{m}.
\end{align}
This implies that for any $\{n,m\}=\{0,1,\dots,l-1\}$,
 \begin{equation}
	\dfrac{d\rho_{n,m}(t)}{d t}=(-i\omega_{n,m}-\gamma_{n,m})\rho_{n,m}(t)+\sum_{k=0}^{l-1}\Gamma_{k\to n}\rho_{k,k}(t)\delta_{n,m}+i\dfrac{{u}(t)}{\hbar}\sum_{k=0}^{l-1}(\mu_{n,k}\rho_{k,m}(t)-\rho_{n,k}(t)\mu_{k,m}),
\end{equation}
where we used the definitions $\omega_{n,m}:=\dfrac{E_n-E_m}{\hbar}$ and $\gamma_{n,m}$ given in Eq.(\ref{gamma_nm}). This proves the form provided in Eq.(\ref{rho_elements}).
 	\section{Vectorisation of the density operator}\label{app_B}
 		\setcounter{equation}{0} 
 	\renewcommand{\theequation}{B.\arabic{equation}}
In this section we provide the vectorisation of the density operator introduced in Section (\ref{Solution}). Although any vectorisation of the density operator can be used, we adopt the tetradic notation introduced in \cite{Mukamel}. The density matrix $\rho(t)$ described in  Eq.(\ref{LVMS}) can be written as follows,
\begin{equation}
	\rho(t)=(\rho(t))^\dagger=\left(
	\begin{array}{ccccccccc}
		\rho_{0,0}(t)&\rho_{0,1}(t) & \dots &\rho_{0,l-1}(t) \\
		\rho_{1,0}(t)&\rho_{1,1}(t) & \dots &\rho_{1,l-1}(t) \\
		\vdots&\vdots & \ddots &\vdots \\
		\rho_{l-1,0}(t)&\rho_{l-1,1}(t) & \dots &\rho_{l-1,l-1}(t) \\
	\end{array}
	\right)\in\mathbf{C}^{l \times l}.\end{equation}
By using the following vectorisation adopted from \cite{Mukamel},
\begin{align}\label{x_t_vet}
	\tilde{x}(t)&=\text{vec}(\rho(t)) \nonumber \\&=\begin{bmatrix}
		\rho_{0,0}(t) & \rho_{1,1}(t) 
		\dots
		\rho_{l-1,l-1}(t)& \rho_{0,1}(t) \dots  
		\rho_{0,l-1}(t)&\rho_{1,0}(t) 
		\dots\rho_{l-1,0}(t) 
		\dots\dots\dots\rho_{l-1,1}(t)\dots\rho_{l-1,l-2}(t) 
	\end{bmatrix}^T,
\end{align}
the differential equations (\ref{rho_elements}) can be equivalently written as,
\begin{equation}\label{NLVN_pa}
	\dfrac{d\tilde{x}(t)}{d t} =(\tilde{A}+iu(t){N})\tilde{x}(t),\hspace{0.5cm} \tilde{x}(0)=\tilde{x}_0,
\end{equation} 
where $\tilde{A} \in\mathbf{C}^{l^2 \times l^2}$,  $\tilde{N} \in\mathbf{C}^{l^2\times l^2}$ are some matrices whose elements can be found from  Eq.(\ref{rho_elements}), and $\tilde{x}_0$ is the vectorisation of the initial density operator \cite{redu}. In addition $T$ in Eq.(\ref{x_t_vet}) stands for the transpose operation. Let $x_e$ be an eigenvector of $\tilde{A}$ with zero eigenvalue, $\tilde{A}x_e=0$ and setting $x(t)=\tilde{x}(t)-x_e$ and $q=Nx_e$, it follows that,
 	\begin{equation}\label{a_til}
 		\dfrac{dx(t)}{d t} =\tilde{A} x(t)+\tilde{B}(x(t)) u(t),\hspace{0.5cm} x(0)={x}_0-x_e,
 	\end{equation}
 	where $\tilde{B}(x(t))=N(x(t)+x_e)=Nx(t)+q$. 

To give an example, using Eq.(\ref{rho_elements}) the matrices $\tilde{A}$ and $N$ appearing in Eq.(\ref{a_til}) for a two dimensional system, i. e. $l=0,1$ can be easily constructed. They are given by,
 	\begin{equation}
 		\tilde{A}=\left(
 		\begin{array}{cccc}
 			-\gamma_{0,0} &\Gamma_{1\to 0}& 0 & 0\\
 			\Gamma_{0\to 1} & -\gamma_{1,1} & 0& 0\\
 			0 & 0 & -i\omega_{0,1}-\gamma_{0,1}& 0 \\
 			0 & 0 & 0& -i\omega_{1,0}-\gamma_{1,0}\\
 		\end{array}\right)
 	\end{equation}
 	and 
 	\begin{equation}
 		{N}=\dfrac{1}{\hbar}\left(
 		\begin{array}{cccc}
 			0 &0& -\mu_{1,0} & \mu_{0,1}\\
 			0 & 0 & \mu_{1,0}& -\mu_{0,1}\\
 			-\mu_{0,1}& \mu_{0,1} & \mu_{0,0}-\mu_{1,1}& 0 \\
 			\mu_{1,0}& -\mu_{1,0} & 0& \mu_{1,1}-\mu_{0,0} \\
 		\end{array}\right).
 	\end{equation}
 	The results can be generalised to any $l>1$ in a straightforward manner. 
 	\section{Complex normal distribution}\label{app_C}
 		\setcounter{equation}{0} 
 	\renewcommand{\theequation}{C.\arabic{equation}}
 For nonsingular covariance matrix  $\Gamma$, the complex normal distribution for a complex random variable $x_t\in \mathbf{C}^n$ is given by,
 \begin{align}\label{norcomplx}
 	&\mathcal{N}_{\mathcal{C}}(\mu_{t},{\Gamma})=\dfrac{1}{\pi^n|\Gamma|} \exp\bigg[- {({{x}_t}- {{\mu} _t})^{\dagger}}{\Gamma}^{-1}({x_t} - {\mu _t})\bigg],
 \end{align}
where $|\Gamma|$ denotes the determinant of $\Gamma$.
 \section{Calculation of the performance index $\gamma(x_{t-1})$}\label{AppD}
 	\setcounter{equation}{0} 
 \renewcommand{\theequation}{D.\arabic{equation}}
In this section, we aim to calculate the form of the performance index  given in Eq. (\ref{gamma_theo}). Let us first evaluate the coefficient $\beta(u_{t-1},x_{t-1})$ defined in  Eq.(\ref{beta}) and repeated here,
 \begin{align}\label{betaapp1}
 	\beta(u_{t-1},x_{t-1}) = \int s(x_{t}|u_{t-1}, x_{t-1}) s(o_t| x_{t}) \ln \bigg ( \frac{s(x_{t}|u_{t-1}, x_{t-1})  s(o_t| x_{t})}{^Is(x_{t}|u_{t-1}, x_{t-1}) ^Is(o_t| x_{t})} \dfrac{1}{ {\gamma}(x_{t})}\bigg) \mathrm{d} x_{t}\mathrm{d} o_t.
 \end{align}
 Let us first calculate  $ \ln \bigg ( \frac{s(x_{t}|u_{t-1}, x_{t-1})  s(o_t| x_{t})}{^Is(x_{t}|u_{t-1}, x_{t-1}) ^Is(o_t| x_{t})} \dfrac{1}{ {\gamma}(x_{t})}\bigg) $. From Eqs.  (\ref{eq:eq4}), (\ref{eq:eq5}),  (\ref{eq:eq1}),  (\ref{eq:eq2}) and (\ref{gamma_theo}), we have,
 \begin{align}\label{betaapp2}
 	&  \ln \bigg ( \frac{s(x_{t}|u_{t-1}, x_{t-1})  s(o_t| x_{t})}{^Is(x_{t}|u_{t-1}, x_{t-1}) ^Is(o_t| x_{t})} \dfrac{1}{ {\gamma}(x_{t})}\bigg)=- {({x_t} - {\mu_t})^\dagger}{\Gamma ^{ - 1}}({x_t} - {\mu_t})+ {({x_t} - {{x}_{r}})^\dagger}\Gamma_r^{ - 1}({x_t}  - {{x}_{r}}) \nonumber\\
 	& - {({o_t} - D{x_t})^\dagger}{G ^{ - 1}}({o_t} - D{x_t})+ {({o_t} - {{o}_{r}})^\dagger}G_r^{ - 1}({o_t} - {{o}_{r}}) +x_{t}^\dagger{M_{t}}{x_{t}} +P_{t}{x_{t}} +\bar{P}_{t}\bar{x}_{t}+ {\omega_{t}}+\ln \bigg( \dfrac{|\Gamma_r|}{|\Gamma|} \dfrac{|G_r|}{|G|}\bigg).\end{align}
 Note that without loss of generality, we absorb the $0.5$ factor in the definition of the standard Gaussian distribution in its covariance matrix. The evaluation of Eq.(\ref{betaapp2}) yields,
 \begin{align}
 	\label{ln}
 	& \ln \bigg ( \frac{s(x_{t}|u_{t-1}, x_{t-1})  s(o_t| x_{t})}{^Is(x_{t}|u_{t-1}, x_{t-1}) ^Is(o_t| x_{t})} \dfrac{1}{ {\gamma}(x_{t})}\bigg)\nonumber\\& =-x_{t}^{\dagger}\Gamma^{-1}x_{t}+x_{t}^{\dagger}\Gamma^{-1}\mu_{t}+\mu_{t}^{\dagger}\Gamma^{-1}x_{t}-\mu_{t}^{\dagger}\Gamma^{-1}\mu_{t}+x_{t}^{\dagger}\Gamma_{r}^{-1}x_{t}-x_{t}^{\dagger}\Gamma_{r}^{-1}x_{r}-x_{r}^{\dagger}\Gamma_{r}^{-1}x_{t}+x_{r}^{\dagger}\Gamma_{r}^{-1}x_{r}\nonumber\\&
 	-o_{t}^{\dagger}G^{-1}o_{t}+o_{t}^{\dagger}G^{-1}D{x}_{t}+x_{t}^{\dagger}D^{\dagger}G^{-1}o_{t}-x_{t}^{\dagger}D^{\dagger}G^{-1}Dx_{t}+o_{t}^{\dagger}G_{r}^{-1}o_{t}-o_{t}^{\dagger}G_{r}^{-1}o_{r}-o_{r}^{\dagger}G_{r}^{-1}o_{t}+o_{r}^{\dagger}G_{r}^{-1}o_{r}\nonumber\\&
 	 +x_{t}^\dagger{M_{t}}{x_{t}} +P_{t}{x_{t}} +\bar{P}_{t}\bar{x}_{t}+ {\omega_{t}+\ln \bigg( \dfrac{|\Gamma_r|}{|\Gamma|} \dfrac{|G_r|}{|G|}\bigg)}.
 \end{align}
 By integrating Eq.(\ref{ln}) over $o_t$, we find,
  \begin{align}
  	\label{ln_2}
 	& \int  s(o_t| x_{t}) \ln \bigg ( \frac{s(x_{t}|u_{t-1}, x_{t-1})  s(o_t| x_{t})}{^Is(x_{t}|u_{t-1}, x_{t-1}) ^Is(o_t| x_{t})} \dfrac{1}{ {\gamma}(x_{t})}\bigg)\text{d}{o}_{t}\nonumber\\& =-x_{t}^{\dagger}\Gamma^{-1}x_{t}+x_{t}^{\dagger}\Gamma^{-1}\mu_{t}+\mu_{t}^{\dagger}\Gamma^{-1}x_{t}-\mu_{t}^{\dagger}\Gamma^{-1}\mu_{t}+x_{t}^{\dagger}\Gamma_{r}^{-1}x_{t}-x_{t}^{\dagger}\Gamma_{r}^{-1}x_{r}-x_{r}^{\dagger}\Gamma_{r}^{-1}x_{t}+x_{r}^{\dagger}\Gamma_{r}^{-1}x_{r}\nonumber\\&
+x_{t}^{\dagger}D^{\dagger}G_{r}^{-1}Dx_{t}-x_{t}^{\dagger}D^{\dagger}G_{r}^{-1}o_{r}-o_{r}^{\dagger}G_{r}^{-1}Dx_{t}+o_{r}^{\dagger}G_{r}^{-1}o_{r}
 	+x_{t}^\dagger{M_{t}}{x_{t}} +P_{t}{x_{t}} +\bar{P}_{t}\bar{x}_{t}+ {\omega_{t}+\ln \bigg( \dfrac{|\Gamma_r|}{|\Gamma|} \dfrac{|G_r|}{|G|}\bigg)}\nonumber\\	& 
 -	\Tr(G({G}^{-1}-{G}_r^{-1})).
 \end{align}
Now, we integrate over $x_t$ to find $\beta$,

 \begin{align}
 	\label{beta_2}
	& 	\beta(u_{t-1},x_{t-1}) = \int s(x_{t}|u_{t-1}, x_{t-1}) s(o_t| x_{t}) \ln \bigg ( \frac{s(x_{t}|u_{t-1}, x_{t-1})  s(o_t| x_{t})}{^Is(x_{t}|u_{t-1}, x_{t-1}) ^Is(o_t| x_{t})} \dfrac{1}{ {\gamma}(x_{t})}\bigg) \mathrm{d} x_{t}\mathrm{d} o_t\nonumber\\& =\mu_{t}^{\dagger}\Gamma_{r}^{-1}\mu_{t}-\mu_{t}^{\dagger}\Gamma_{r}^{-1}x_{r}-x_{r}^{\dagger}\Gamma_{r}^{-1}\mu_{t}+x_{r}^{\dagger}\Gamma_{r}^{-1}x_{r}
	+\mu_{t}^{\dagger}D^{\dagger}G_{r}^{-1}D\mu_{t}-\mu_{t}^{\dagger}D^{\dagger}G_{r}^{-1}o_{r}-o_{r}^{\dagger}G_{r}^{-1}D\mu_{t}+o_{r}^{\dagger}G_{r}^{-1}o_{r}
	+\mu_{t}^\dagger{M_{t}}{\mu_{t}}\nonumber\\& +P_{t}{\mu_{t}} +\bar{P}_{t}\bar{\mu}_{t}+ {\omega_{t}+\ln \bigg( \dfrac{|\Gamma_r|}{|\Gamma|} \dfrac{|G_r|}{|G|}\bigg)}-\Tr(G({G}^{-1}-{G}_r^{-1}))-	\Tr(\Gamma({\Gamma}^{-1}-{\Gamma}_r^{-1}-D^{\dagger}G_r^{-1}{D}-M_t))
	\nonumber\\& =\mu_{t}^\dagger(\Gamma_r^{-1}+D^{\dagger}G_r^{-1}D+M_{t})\mu_{t}-\mu_{t}^\dagger(\Gamma_r^{-1}x_{r}+D^{\dagger}G_r^{-1}o_r-P_{t}^{\dagger})-(x_r^{\dagger}\Gamma_r^{-1}+o_r^\dagger{G}_r^{-1}D-P_t)\mu_t+x_r^{\dagger}\Gamma_r^{-1}{x}_r+o_r^{\dagger}G_r^{-1}o_r+\omega_{t}	\nonumber\\&+\ln \bigg( \dfrac{|\Gamma_r|}{|\Gamma|} \dfrac{|G_r|}{|G|}\bigg)-\Tr(G({G}^{-1}-{G}_r^{-1}))-	\Tr(\Gamma({\Gamma}^{-1}-{\Gamma}_r^{-1}-D^{\dagger}G_r^{-1}{D}-M_t)).
\end{align}
Setting,
 \begin{align}
	C=& \Gamma_r^{-1}+D^{\dagger}G_r^{-1}D+M_{t},
\nonumber\\
  J=	&x_r^{\dagger}\Gamma_r^{-1}+o_r^\dagger{G}_r^{-1}D-P_t,
  \nonumber\\ I=&
  \ln \bigg( \dfrac{|\Gamma_r|}{|\Gamma|} \dfrac{|G_r|}{|G|}\bigg)-\Tr(G({G}^{-1}-{G}_r^{-1}))-	\Tr(\Gamma({\Gamma}^{-1}-{\Gamma}_r^{-1}-D^{\dagger}G_r^{-1}{D}-M_t)),
\end{align}
and remembering that $\mu_t=Ax_{t-1}+Bu_{t-1}$, the form given in Eq.(\ref{beta_2}) can be simplified as follows,
\begin{align}
	\label{betaapp5}
	& 	\beta(u_{t-1},x_{t-1}) = x_{t-1}^\dagger{A}^{\dagger}C{A}{x}_{t-1}+x_{t-1}^\dagger{A}^{\dagger}C{B}{u}_{t-1}+u_{t-1}^\dagger{B}^{\dagger}C{A}{x}_{t-1}+u_{t-1}^\dagger{B}^{\dagger}C{B}{u}_{t-1}-JAx_{t-1}-JBu_{t-1}-\bar{J}\bar{A}\bar{x}_{t-1}\nonumber\\&-\bar{J}\bar{B}\bar{u}_{t-1}+(x_r^{\dagger}\Gamma_r^{-1}{x}_r+o_r^{\dagger}G_r^{-1}o_r+\omega_{t}+I).
\end{align}
By projecting the form of $\beta(u_{t-1},x_{t-1})$ found in Eq.(\ref{betaapp5}) along with the ideal distribution of the controller provided in  Eq.(\ref{eq:eq3}), in the definition of $\gamma(x_{t-1})$ given in  Eq.(\ref{gamma2}), we find that,
\begin{align}
	& 	\gamma(x_{t-1}) = \int {^Ic(u_{t-1}|x_{t-1}) \exp\big({-\beta(u_{t-1},x_{t-1})}\big) \mathrm{d} u_{t-1}}= \dfrac{1}{\pi|\Omega|}\int  \exp\bigg(- {(u_{t-1} - {u_r})^\dagger}{\Omega ^{ - 1}}({u_{t-1}} - {u_r})-x_{t-1}^\dagger{A}^{\dagger}C{A}{x}_{t-1}\nonumber\\&-x_{t-1}^\dagger{A}^{\dagger}C{B}{u}_{t-1}-u_{t-1}^\dagger{B}^{\dagger}C{A}{x}_{t-1}-u_{t-1}^\dagger{B}^{\dagger}C{B}{u}_{t-1}+JAx_{t-1}+JBu_{t-1}+\bar{J}\bar{A}\bar{x}_{t-1}+\bar{J}\bar{B}\bar{u}_{t-1}-(x_r^{\dagger}\Gamma_r^{-1}{x}_r+o_r^{\dagger}G_r^{-1}o_r\nonumber\\&+\omega_{t}+I)\bigg) \mathrm{d} u_{t-1}
\end{align}
Simplifying the above equation by collecting quadratic, linear and constant terms with respect to the control signal, $u_{t-1}$ together, yields,
\begin{align}
	& 	\gamma(x_{t-1})=\dfrac{1}{\pi|\Omega|}\int  \exp\bigg(- u_{t-1}^\dagger\big({\Omega^{-1}}+B^\dagger{C}B\big)u_{t-1}-u_{t-1}^\dagger\big(B^\dagger{C}Ax_{t-1}-\Omega^{-1}u_{r}-B^\dagger{J}^\dagger\big) \nonumber \\ &-u_{t-1}^T\big(B^TC^T\bar{A}\bar{x}_{t-1}-(\Omega^{-1})^T\bar{u}_r -B^TJ^T\big)-\big(u_r^\dagger\Omega^{-1}u_r+x_{t-1}^\dagger{A}^{\dagger}C{A}{x}_{t-1}-JAx_{t-1}-\bar{J}\bar{A}\bar{x}_{t-1} \nonumber\\&+x_r^{\dagger}\Gamma_r^{-1}{x}_r+o_r^{\dagger}G_r^{-1}o_r+\omega_{t}+I\big)  \bigg)\mathrm{d}u_{t-1}.
\end{align}
Introduce the following definitions,
\begin{align}
	\label{defi}
	& F_t={\Omega^{-1}}+B^\dagger{C}B,
	\nonumber\\&Q_t^\dagger=B^\dagger{C}Ax_{t-1}-\Omega^{-1}u_{r}-B^\dagger{J}^\dagger,
	\nonumber\\&\text{and}\nonumber\\&
	v_{t-1}=-F_{t}^{-1}Q_t^\dagger.
\end{align}
It thus follows that,
\begin{align}
	\label{suite_gamma}
	& 	\gamma(x_{t-1})=\dfrac{1}{\pi|\Omega|}\int  \exp\bigg( {-({u_{t-1}} - {v_{t-1}})^\dagger}{F}_{t}({u_{t-1}} - {v_{t-1}})\bigg)\mathrm{d}u_{t-1}\nonumber\\	&  \exp\bigg[-\bigg(-v_{t-1}^\dagger{F_t}v_{t-1}+x_{t-1}^\dagger{A}^{\dagger}C{A}{x}_{t-1}-JAx_{t-1}-\bar{J}\bar{A}\bar{x}_{t-1}+u_r^\dagger\Omega^{-1}u_r+x_r^{\dagger}\Gamma_r^{-1}{x}_r+o_r^{\dagger}G_r^{-1}o_r+\omega_{t}+I\bigg)\bigg].
\end{align}
The first exponential function in Eq.(\ref{suite_gamma}) is nothing but the complex normal distribution recalled in Eq.(\ref{norcomplx}) which when integrated over $u_{t-1}$ yields a normalisation constant.
Hence, 
\begin{align}
	\label{gamm_app}
	& 	\gamma(x_{t-1})=\dfrac{|F_{t}^{-1}|}{|\Omega|} \exp\bigg[-\bigg(-Q^T_{t}{F_t}^{-1}Q_{t}+u_r^\dagger\Omega^{-1}u_r+x_{t-1}^\dagger{A}^{\dagger}C{A}{x}_{t-1}-JAx_{t-1}-\bar{J}\bar{A}\bar{x}_{t-1}+x_r^{\dagger}\Gamma_r^{-1}{x}_r+o_r^{\dagger}G_r^{-1}o_r+\omega_{t}\nonumber\\& +I\bigg)\bigg],
\end{align}
Taking $-\ln$ of both sides of the above equation yields,
	\begin{align}
	- \ln \left( {\gamma \left( {{x_{t-1}}} \right)} \right) &=x_{t-1}^{\dagger}A^{\dagger}\bigg(C-CBF_{t}^{-1}B^\dagger{C}\bigg)Ax_{t-1}+\bigg((u_r^\dagger\Omega^{-1}+JB)F_{t}^{-1}B^{\dagger}C-J\bigg)Ax_{t-1}
	\nonumber\\&+\bigg((u_r^T\Omega^{-1}+\bar{J}\bar{B})F_{t}^{-1}B^{T}\bar{C}-\bar{J}\bigg)\bar{A}\bar{x}_{t-1}+u_r^\dagger\Omega^{-1}u_r+x_r^{\dagger}\Gamma_r^{-1}{x}_r+o_r^{\dagger}G_r^{-1}o_r+\omega_{t}+I- \ln \left(\dfrac{|F_{t}^{-1}|}{|\Omega|} \right).
	\end{align}
This completes the proof of the Theorem (\ref{Theo1}).
	\section{Distribution of the optimal control }\label{AppE}
\setcounter{equation}{0} 
\renewcommand{\theequation}{E.\arabic{equation}}
In this section we provide the form of 	the distribution of the optimal control that  minimises the Kullback-Leibler distance between (\ref{gpdf2}) and (\ref{igpdf2}). It is given in Eq.(\ref{eq:Eqn4}), repeated here,
\begin{equation}
	\label{id_app}
	c(u_{t-1}|x_{t-1})=\frac{^Ic(u_{t-1}|x_{t-1}) \exp [-\beta(u_{t-1},x_{t-1})]}{\gamma(x_{t-1})}.
\end{equation} 
By substituting the pdf of the ideal controller $^Ic(u_{t-1}|x_{t-1})$, the coefficient $\beta(u_{t-1},x_{t-1})$ and the performance index $ \gamma \left( {{x_{t-1}}} \right)$, given in Eqs.(\ref{eq:eq3}), (\ref{beta_2}) and (\ref{gamm_app}), respectively in Eq.(\ref{id_app}), we get,
\begin{align}
 &c(u_{t-1}|x_{t-1})=\dfrac{1}{\pi|\Omega|}\exp\bigg[- {u_t}^\dagger\big({\Omega^{-1}}+B^\dagger{C}B\big){u_t}-{u_t}^\dagger\big(B^\dagger{C}Ax_{t-1}-\Omega^{-1}u_{r}-B^\dagger{J}^\dagger\big)-u_{t-1}^T\big(B^TC^T\bar{A}\bar{x}_{t-1}-(\Omega^{-1})^T\bar{u}_r\nonumber\\&-B^TJ^T\big)-\big(u_r^\dagger\Omega^{-1}u_r+x_{t-1}^\dagger{A}^{\dagger}C{A}{x}_{t-1}-JAx_{t-1}-\bar{J}\bar{A}\bar{x}_{t-1}+x_r^{\dagger}\Gamma_r^{-1}{x}_r+o_r^{\dagger}G_r^{-1}o_r+\omega_{t}+I\big) \nonumber\\&
+x_{t-1}^{\dagger}A^{\dagger}\bigg(C-CBF_{t}^{-1}B^\dagger{C}\bigg)Ax_{t-1}+\bigg((u_r^\dagger\Omega^{-1}+JB)F_{t}^{-1}B^{\dagger}C-J\bigg)Ax_{t-1}
 +\bigg((u_r^T\Omega^{-1}+\bar{J}\bar{B})F_{t}^{-1}B^{T}\bar{C}-\bar{J}\bigg)\bar{A}\bar{x}_{t-1}\nonumber\\&+u_r^\dagger\Omega^{-1}u_r+x_r^{\dagger}\Gamma_r^{-1}{x}_r+o_r^{\dagger}G_r^{-1}o_r+\omega_{t}+I- \ln \left(\dfrac{|F_{t}^{-1}|}{|\Omega|} \right)\bigg],
\end{align}
after simplification we find, 
\begin{equation}
	c(u_{t-1}|x_{t-1})=\dfrac{1}{\pi|F_{t}^{-1}|}\exp\big(- ({u_t}-{v_t})^\dagger{F_t}({u_t}-{v_t})\big),
\end{equation}
where $v_t$ and $F_t$ are defined in Eq.(\ref{defi}). This means that,
\begin{equation}
	c(u_{t-1}|x_{t-1})\sim \mathcal{N}(v_{t-1},F_{t}^{-1}).
\end{equation}
	\section{State space model for spin $\frac{1}{2}$}\label{spinn_1_2}
\setcounter{equation}{0} 
\renewcommand{\theequation}{F.\arabic{equation}}
In this section we show the forms of the matrices $\tilde{A}$ and $\tilde{N}$ appearing in Eq.(\ref{NLVN}) for a spin $\dfrac{1}{2}$ interacting only with an external electric field. We assume that this interaction can be described by the following master equation,
\begin{equation}\label{vonspin}
	\dfrac{d\rho(t)}{dt}=-i[H,\rho(t)],
\end{equation}
where as already stated in  Eq.(\ref{58_}), $H=\dfrac{1}{2}\sigma_3+\dfrac{1}{2}(\sigma_1+\sigma_2) u(t)$, is defined in terms of the electric field $u(t)$  and the Pauli matrices  $\sigma_1,\sigma_2, \sigma_3$ are as given in  Eq.(\ref{PauliMat}), repeated here, 
\begin{equation}
	\sigma_1	=\left(
	\begin{array}{cc}
		0&1\\
		1 & 0 \\
	\end{array}
	\right),\hspace{0.4cm}	\sigma_2=\left(
	\begin{array}{cc}
		0 &-i\\
		i & 0 \\
	\end{array}
	\right),\hspace{0.4cm}\sigma_3	=\left(
	\begin{array}{cc}
		1 &0\\
		0 & -1\\
	\end{array}
	\right).
\end{equation}
The matrix representation of the Hamiltonian, $H$ can be then easily evaluated. It is given by,
\begin{equation}
	H=\dfrac{1}{2}\left(
	\begin{array}{cc}
		1 & u(t)(1-i) \\
		u(t)(1+i) & -1 \\
	\end{array}
	\right).
\end{equation} 
By considering,
\begin{equation}
	\rho(t)=\left(
	\begin{array}{cc}
		\rho_{00}(t) & \rho_{01}(t)\\
		\rho^*_{01}(t)  & \rho_{11}(t)
	\end{array}
	\right),
\end{equation} 
the Liouville von-Neumann equation (\ref{vonspin}) can be re-written as,
\begin{equation}
	\dfrac{d}{dt}\left(
	\begin{array}{cc}
		\rho_{00}(t) & \rho_{01}(t)\\
		\rho^*_{01}(t)  &\rho_{11}(t)
	\end{array}
	\right)=\dfrac{-i}{2}\left(
	\begin{array}{cc}
		u(t)\big((1-i) \rho^*_{01}(t)-(1+i) \rho_{01}(t)\big) &2\rho_{01}(t)+u(t)(1-i)(\rho_{11}(t)-\rho_{00}(t))\\
		-2\rho^*_{01}(t) +u(t)(1+i)(\rho_{00}(t) -\rho_{11}(t) ) & u(t)\big((1+i)\rho_{01}(t) -(1-i)\rho^*_{01}(t) \big)  \\
	\end{array}
	\right).
\end{equation}
Using the vectorisation defined in Eq.(\ref{x_t_vet}), it follows that,
\begin{align}
	&	\dfrac{d}{dt}\underbrace{\left(
		\begin{array}{cc}
			\rho_{00}(t)\\ \rho_{11}(t)\\
			\rho_{01}(t) \\ \rho^*_{01}(t)
		\end{array}
		\right)}_{x(t)}= \underbrace{\left(
		\begin{array}{cccc}
			0 & 0 & 0 & 0\\
			0 & 0 & 0& 0\\
			0 & 0 & -i& 0 \\
			0 & 0 & 0& i\\
		\end{array}\right)}_{\tilde{A}} \underbrace{\left(
		\begin{array}{cc}
			\rho_{00}(t)\\ \rho_{11}(t)\\
			\rho_{01}(t) \\ \rho^*_{01}(t)
		\end{array}
		\right)}_{x(t)}\\&+i \underbrace{\dfrac{ 1}{2}\left(
		\begin{array}{cccc}
			0 & 0 & (1+i) & -(1-i)\\
			0 & 0 & -(1+i) & (1-i)\\
			(1-i) & -(1-i)&0 & 0 \\
			-(1+i) & (1+i)&0 & 0 \\
		\end{array}\right)}_{\tilde{N}} \underbrace{\left(
		\begin{array}{cc}
			\rho_{00}(t)\\ \rho_{11}(t)\\
			\rho_{01}(t) \\ \rho^*_{01}(t)
		\end{array}
		\right)}_{x(t)}u(t),
\end{align}
yielding the state equation for the spin-1/2 system in the form given in  Eq.(\ref{NLVN_pa}), repeated here,
\begin{equation}
	\dfrac{d{x}(t)}{d t} =(\tilde{A}+i\tilde{N} u(t) ){x}(t).
\end{equation}
\section{State space model for spin-1}\label{spinn_1}
\setcounter{equation}{0} 
\renewcommand{\theequation}{G.\arabic{equation}}
Similar to spin $\dfrac{1}{2}$, we aim in this section to find the state equation describing the interaction of a spin $1$ system with an electric field $u(t)$. Let us consider, without loss of generality, that the Hamiltonian describing the interaction between a spin-1 system and an electric field $u(t)$ is given by, 
\begin{equation}
	H=H_0+H_u(t)=\left(\begin{array}{ccc}
		\dfrac{3}{2} & 0 & 0  \\
		0 & 1 & 0  \\
		0 & 0&0   \\
	\end{array}\right)+\left(\begin{array}{ccc}
		0 & 0 & 1  \\
		0 & 0 & 1  \\
		1 & 1&0   \\
	\end{array}\right) u(t)=\left(\begin{array}{ccc}
		\dfrac{3}{2} & 0 & u(t)  \\
		0 & 1& u(t)  \\
		u(t) & u(t)&0   \\
	\end{array}\right)
\end{equation} 
Setting,
\begin{equation}
	\rho=\left(\begin{array}{ccc}
		\rho_{00}(t) & 	\rho_{01}(t) & 	\rho_{02}(t)\  \\
		\rho^*_{01}(t) & 	\rho_{11}(t) & \rho_{12}(t)\ \\
		\rho^*_{02}(t) &  \rho^*_{12}(t) &	\rho_{22}(t)
	\end{array}\right),
\end{equation}
the master equation $	\dfrac{d\rho(t)}{dt}=-i[H,\rho(t)]$ can be written as, 
{\small	\begin{align}
		&	i\dfrac{d}{dt}\left(\begin{array}{ccc}
			\rho_{00}(t) & 	\rho_{01}(t) & 	\rho_{02}(t)\  \\
			\rho^*_{01}(t) & 	\rho_{11}(t) & \rho_{12}(t)\ \\
			\rho^*_{02}(t) &  \rho^*_{12}(t) &	\rho_{22}(t)
		\end{array}\right)=\\&\left(\begin{array}{ccc}
			u(t)\big(\rho^*_{02}(t) -\rho_{02}(t)\big)& \dfrac{1}{2}\rho_{01}(t)+u(t)\big( \rho^*_{12}(t)-\rho_{02}(t)\big ) & \dfrac{3}{2}\rho_{02}(t) +u(t)\big(\rho_{22}(t)-\rho_{00}(t) -\rho_{01}(t)\big)  \\\\
			-\dfrac{1}{2} \rho^*_{01}(t)+u(t)(\rho^*_{02}(t)- \rho_{12}(t))& u(t)\big( \rho_{12}^*(t)- \rho_{12}(t)\big)& \rho_{12}(t)+u(t)\big(\rho_{22}(t) -\rho_{01}^*(t)-\rho_{11}(t)\big) \\\\
			-\dfrac{3}{2} \rho_{02}^*(t) +u(t)\big(\rho_{00} +\rho_{01}^*(t)-\rho_{22}(t)\big)& - \rho_{12}^*(t)+u(t)(\rho_{01}(t)+\rho_{11}(t)-\rho_{22} (t)) & u(t)\big(\rho_{02}(t) + \rho_{12}(t)-\rho_{02}^*(t) - \rho_{12}^*(t)\big) \\
		\end{array}\right).
\end{align}}
which when using the vectorisation defined in Eq.(\ref{x_t_vet}), gives,
\newpage
\begin{eqnarray}
	&\dfrac{d}{dt}\underbrace{\left(
		\begin{array}{ccccccccc}
			\rho_{00} (t) \\
			\rho_{11} (t)\\
			\rho_{22}(t) \\
			\rho_{01}(t) \\
			\rho_{02}(t)  \\
			\rho_{01} ^* (t) \\
			\rho_{02}^*(t)  \\
			\rho_{12}(t)\\
			\rho_{12}^* (t) \\
		\end{array}\right)}_{x(t)}=\underbrace{\left(
		\begin{array}{ccccccccc}
			0 & 0 & 0 & 0 & 0 & 0 & 0 & 0 & 0 \\
			0 & 0 & 0 & 0 & 0 & 0 & 0 & 0 & 0 \\
			0 & 0 & 0 & 0 & 0 & 0 & 0 & 0 & 0 \\
			0 & 0 & 0 & -\dfrac{i}{2} & 0 & 0 & 0 & 0 & 0 \\
			0 & 0 & 0 & 0 & -\dfrac{3i}{2}& 0 & 0 & 0 & 0 \\
			0 & 0 & 0 & 0 & 0 & \dfrac{i}{2} & 0 & 0 & 0 \\
			0 & 0 & 0 & 0 & 0 & 0 & \dfrac{3i}{2} & 0 & 0 \\
			0 & 0 & 0 & 0 & 0 & 0 & 0 & -i& 0 \\
			0 & 0 & 0 & 0 & 0 & 0 & 0 & 0 & i\\
		\end{array}
		\right)}_{\tilde{A}} \underbrace{\left(
		\begin{array}{ccccccccc}
			\rho_{00} (t) \\
			\rho_{11} (t)\\
			\rho_{22}(t) \\
			\rho_{01}(t) \\
			\rho_{02}(t)  \\
			\rho_{01} ^* (t) \\
			\rho_{02}^*(t)  \\
			\rho_{12}(t)\\
			\rho_{12}^* (t) \\
		\end{array}\right)}_{x(t)} \nonumber \\
	&+i\underbrace{\left(
		\begin{array}{ccccccccc}
			0 & 0 & 0 & 0 & 1 & 0 & -1 & 0 & 0 \\
			0 & 0 & 0 & 0 & 0 & 0 & 0 & 1 & -1 \\
			0 & 0 & 0 & 0 & -1& 0 & 1 & -1 & 1 \\
			0 & 0 & 0 & 0 & 1 & 0 & 0 & 0 & -1 \\
			1 & 0 & -1 & 1 & 0 & 0 & 0 & 0 & 0 \\
			0 & 0 & 0 & 0 & 0 & 0 & -1 & 1 & 0 \\
			-1 & 0 & 1& 0 & 0 & -1 & 0 & 0 & 0 \\
			0 & 1 & -1 & 0 & 0 & 1 & 0 & 0 & 0 \\
			0 & -1 & 1 & -1 & 0 & 0 & 0 & 0 & 0 \\
		\end{array}
		\right)}_{\tilde{N}}\underbrace{\left(
		\begin{array}{ccccccccc}
			\rho_{00} (t) \\
			\rho_{11} (t)\\
			\rho_{22}(t) \\
			\rho_{01}(t) \\
			\rho_{02}(t)  \\
			\rho_{01} ^* (t) \\
			\rho_{02}^*(t)  \\
			\rho_{12}(t)\\
			\rho_{12}^* (t) \\
		\end{array}\right)}_{x(t)} u(t).
\end{eqnarray}
Hence, we find the form of the state equation  given in Eq.(\ref{NLVN_pa}) as follows,
\begin{equation}
	\dfrac{d{x}(t)}{d t} =(\tilde{A}+i\tilde{N} u(t) ){x}(t).
\end{equation}
	
\newpage{\pagestyle{empty}\cleardoublepage}

\end{document}